\documentclass[preprint,5p,authoryear]{elsarticle}
\usepackage{graphicx}
\usepackage{epstopdf, epsfig}

\usepackage{mathtools}
\usepackage{amsmath}
\usepackage{amssymb}

\begin{document}

%\shorttitle{Lagrangian velocity gradients in LES}
%\shortauthor{P. L. Johnson and P. Moin}

\title{Predicting the impact of particle-particle collisions on turbophoresis with a reduced number of computational particles}

\author{Perry L. Johnson\corref{cor1}}
\ead{perryj@stanford.edu}
\cortext[cor1]{Corresponding author}
%
%	\fnref{fn1}}
%\author{{\color{red}(Parviz Moin?)}}
%\ead{moin@stanford.edu}

\address{Center for Turbulence Research, Stanford University,	Stanford, CA 94305, USA}

\begin{abstract}
	A common feature of wall-bounded turbulent particle-laden flows is enhanced particle concentrations in a thin layer near the wall due to a phenomenon known as turbophoresis. Even at relatively low bulk volume fractions, particle-particle collisions regulate turbophoresis in a critical way, making simulations sensitive to collisional effects. Lagrangian tracking of every particle in the flow can become computationally expensive when the physical number of particles in the system is large. Artificially reducing the number of particles in the simulation can mitigate the computational cost. When particle-particle collisions are an important aspect determining the simulation outcome, as in the case when turbophoresis plays an active role, simply reducing the number of particles in the simulation significantly alters the computed particle statistics. This paper introduces a computational particle treatment for particle-particle collisions which reproduces the results of a full simulation with a reduced number of particles. This is accomplished by artificially enhancing the particle collision radius based on scaling laws for the collision rates. The proposed method retains the use of deterministic collision models and is applicable for both low and high Stokes number regimes.
\end{abstract}

\begin{keyword}
	particle-laden flows \sep Eulerian-Lagrangian \sep computational particles \sep four-way coupling \sep wall-bounded turbulence 
\end{keyword}

\maketitle

\section{Introduction}

Encounters with particle-laden flows are common in both engineering and natural sciences. 
%While the governing equations for simulating fluid flows with dispersed particles, the wide range of length and time scales often involved in applications prohibits a direct approach, which requires sub-particle scale resolution to capture the particle-fluid boundary and apply the relevant boundary or interface equations \citep{Burton2005, Ardekani2016, Sierakowski2016, Horne2019}. 
Because of the wide range of length and time scales typically involved in fluid flows with dispersed particles, it is often infeasible to directly simulate these flows with the sub-particle scale resolution necessary to capture the particle-fluid interface and apply the relevant boundary or interface conditions \citep{Burton2005, Ardekani2016, Sierakowski2016, Horne2019}. On the other hand, treating the collection of particles as a continuum provides a vast reduction in computational cost, but such an approximation may not always be appropriate for accurate simulations \citep{Druzhinin1998, Fox2008}. In between these two approaches, the popular Eulerian-Lagrangian method employs a `point-particle' or `discrete element' description of individual particles as they are tracked along their trajectories through the flow simulated on fixed grid \citep{Squires1990, Capecelatro2013, Subramaniam2013}. Each particle is associated with a number of computational degrees of freedom describing its location, velocity, and other relevant properties. The interaction of a particle with the surrounding flow or other particles is not resolved (by the Eulerian grid), so closures are needed in the form of drag laws, collision models, and other dynamical relationships for integrating ordinary differential equations describing the evolution of the particle \citep{Maxey1983, Gatignol1983}. The computational cost grows (at least) linearly with the number of particles, and can grow faster in cases where identifying and computing particle-particle collisions is important.

Given that systems with billions, trillions, and more particles are not uncommon, Eulerian-Lagrangian approach to particle-laden flows can be inconveniently, or even prohibitively, expensive. A reduction in cost may be achieved by having a single computational particle `represent' many particles in the physical system \citep{Elghobashi1993, Elghobashi1994}. Generally, the goal of such methods to have the statistics of the computational particles accurately reproduce the particle statistics of the case where the physical number of particles is simulated. \citet{Garg2009} described this as statistical equivalence and proscribed a method for dynamically assigning statistical weights to computational particles. In the absence of two-way coupling or particle-particle collisions, a minimal constraint  is that a sufficient number of particles should be used to obtain enough samples for computing statistical quantities of interest \citep{Elghobashi1992, Elghobashi1993}. \cite{Garg2009} used evolving statistical weights with annihilation and cloning of computational particles to reduce the non-uniformity in the computational particle number density, thus making statistical errors relatively uniform in space. This scheme also helps with load balancing for parallelization across many processors. The physics-based necessity of including two-way coupling at high mass fractions and four-way coupling at high volume fractions leads to additional complexities. The two-way coupling force for computational particles must be amplified in proportion to the number of physical particles represented \citep{Garg2009}. This can exacerbate the induced fluid velocity at the particle location and highlight the need for estimating an undisturbed flud velocity \citep{Horwitz2016, Horwitz2018, Horwitz2019}. \citet{Moin2006} presented a scheme to prevent the number of computational particles from exceeding a specified threshold during liquid jet breakup, without considering the impact on collisions or coalescence. In this work, the treatment of collisional effects is isolated for study.

The present work is motivated by the impact of turbophoresis and non-uniform particle concentrations on radiation transport and absorption in internal wall-bounded turbulent flows \citep{Frankel2017, Banko2018}. Turbophoresis is a phenomenon in which small, heavy particles experience a net drift from regions of high turbulence intensity toward lower intensity regions in a turbulent shear flow \citep{Caporaloni1975, Reeks1983, Guha2008}. In wall-bounded flows, the fluid velocity vanishes at the wall due to the no-slip and no-penetration boundary conditions. Within a thin layer of flow adjacent to the wall, therefore, the turbulent fluctuations are less intense. Particles in this region near the wall tend to have lower velocity fluctuations themselves compared with the bulk of the flow away from the wall, leading to a differential in dispersion rates depending on distance from the boundary. Particles are rapidly dispersed throughout most of the turbulent flow but are `trapped' near the wall for longer durations. This results in higher particle concentrations in a thin layer adjacent to the wall and interactions between particles and near wall coherent structures become important \citep{Marchioli2002, Sikovsky2014}.

Particle-particle collisions have a strong impact on turbophoresis, signficantly reducing the non-uniformity of particle concentration profiles as the volume fraction is increased \citep{Li2001}. The increase in particle concentration levels leads to higher collision rates in the near wall region \citep{Kuerten2016}. Particles in low-speed and high-speed streaks experience elevated relative velocities. Oblique collisions redistribute particle streamwise velocity fluctuations into the wall-normal direction \citep{Caraman2003}. Thus, the wall-normal mixing of particles near the wall is greatly enhanced, flattening the concentration profiles even at relatively low bulk volume fractions $\sim 10^{-4}$ \citep{Yamamoto2001, Kuerten2015}. The quadratic scaling of collision rate with particle number density creates the scenario in which the particle concentration profile, as a function of wall-normal distance, is very sensitive to the bulk volume fraction of the particle-laden flow \citep{Johnson2018, Johnson2019}. Any approach which relies on reducing computational cost by decreasing the number of simulated particles will necessarily struggle to provide an accurate concentration field without additional treatment of particle-particle collisions to account for this effect. Because these collisions occur in a very thin layer near the wall, where the flow and particle motions are highly anisotropic, and the obliqueness of collisions is important, stochastic approaches for collisions are less promising.

The rate at which collisions between two particles occur scales quadratically with the local number density of particles \citep{Wang2000, Reade2000}, so simply reducing the number of particles and assigning a statistical weight to each particle leads to under-prediction of collisional effects. Others approaches have considered stochastic models for particle-particle collisions, but this increases the difficulty and imprecision of the collision treatment \citep{Schmidt2000, ORourke2009}. \citet{Sommerfeld2001} introduced an approach based on the creation of fictitious collision partners with random obliqueness  and relative velocity specified by observed trends in homogeneous turbulence. It is not clear how universal such a treatment would be, particularly in highly anisotropic and inhomogeneous regions near the wall in a turbulent flow.

In this paper, a novel method is introduced to address particle-particle collisions within the general computational particle framework of \citet{Garg2009}. The deterministic treatment of collisions is retained. The method consists in artificially enhancing the collisional radius of the particles so as to maintain accurate collision rates as the number of computational particles is reduced. The collisional radius determining the instance and outcome of a collision between two particles. Scaling arguments for the collision rate can be used to determine how the altered collision radius depends on its statistical weight. The scaling may be directly determined in the opposite limits of low and high Stokes numbers, and a hybrid model is proposed for bridging between these limits. The resulting approach dynamically computes the collision radius for each individual collision based on the properties of the particle and fluid involved.

The rest of the paper is organized as follows. The sensitivity of the particle concentration field to particle-particle collisions in a canonical flow is considered in detail in \S \ref{sec:background}. In \S \ref{sec:theory}, the theoretical basis for reducing the number of computational particles is reviewed and extended to include collisional effects by introducing scaling relations for an enhanced collisional radius treatment. Simulation results in \S \ref{sec:results} for particle-laden turbulent channel flow demonstrate that such scaling arguments work in the low and high Stokes number limits. Further, a hybrid model bridging these limits is constructed and demonstrated to work well in both regimes. Finally, \S \ref{sec:conclusion} provides a summary and concluding discussion.

\section{Turbophoresis and particle-particle collisions \label{sec:background}}
%Turbophoresis is the process by which turbulent fluid fluctuations cause a net drift of inertial particles from regions of high turbulence intensity toward lower turbulence intensity \citep{Caporaloni1975, Reeks1983}.
This section demonstrates the difficulties associated with reducing the number of computational particles in wall-bounded turbulent flow simulations. The basic physics of turbophoresis are reviewed (\S \ref{sec:turbophoresis}) in a statistical framework (\S \ref{sec:statistics}) and the impact of particle-particle collisions is demonstrated (\S \ref{sec:collisions}), with particular attention to the particle concentration profile near the wall. To set the stage, a fully-developed particle-laden turbulent channel flow is introduced in \S \ref{sec:channel-numerical} as the canonical flow for the current study. The physical model and numerical approach, which combines a Lagrangian description of the particle dynamics with direct numerical simulation (DNS) of the fluid on an Eulerian grid, are also described in \S \ref{sec:channel-numerical}.

\subsection{Canonical flow: particle-laden turbulent channel \label{sec:channel-numerical}}
The turbulent channel flow is adopted to demonstrate the basic physical principles of turbophoresis and the impact of particle-particle collisions. The flow is bounded only in the $y$ (wall-normal) direction by two smooth, flat walls separated by a distance $2h$, where $h$ denotes the half-height of the channel. A pressure gradient, $-dp/dx = \rho_f u_*^2 / h$, forces the mean flow in the $x$ (streamwise) direction. The statistics of the turbulent flow and the particles embedded in the flow are homogeneous with respect to the $x$ and $z$ (spanwise) directions. Here, $\rho_f$ is the fluid mass density and $u_*$ is the friction velocity which characterize the typical turbulent velocity fluctuation magnitude. The friction Reynolds number, $Re_* = u_* h / \nu_f$, where  $\nu_f$ is the kinematic viscosity of the fluid, characterizes the range of active turbulent length scales between the geometry-imposed length scale, $h$, and viscous length scale, $\delta_* = \nu_f / u_*$. It also encapulates the range of active timescales in the turbulent fluctuations, from the large-eddy turnover time, $h / u_*$, to the viscous timescale, $\tau_* = \nu_f / u_*^2$.

The numerical simulations used in this work employ periodic boundary conditions with domain size $L_x = 4\pi h$ and $L_z = 2\pi h$. The flow is assumed incompressible and is described by a velocity field, $\mathbf{u}(\mathbf{x},t)$, and a pressure field, $p(\mathbf{x},t)$, which evolve according to,
\begin{equation}
\frac{\partial \mathbf{u}}{\partial t} + \mathbf{u} \cdot \nabla \mathbf{u} = -\frac{1}{\rho_f}\nabla p + \nu_f \nabla^2 \mathbf{u} + \mathbf{f}, \hspace{0.04\linewidth} \nabla \cdot \mathbf{u} = 0,
\label{eq:Navier-Stokes}
\end{equation}
where and $f_i$ represents the mean pressure gradient forcing as well as any two-way coupling force from the particles. Numerical simulations for this paper solve Eq. \eqref{eq:Navier-Stokes} on a staggered Cartesian grid with uniform spacing $\Delta x^+ = \Delta x / \delta_* \approx 11$, $\Delta z^+ = \Delta z / \delta_* \approx 7$ in the streamwise and spanwise directions, respectively. The grid is non-uniform in the wall-normal direction with $\Delta y_{min}^+ \approx 0.5$ adjacent to the wall and $\Delta y_{max}^+ \approx 7$ at the center of the channel. The discretization of spatial derivatives uses second-order central differencing and a second-order Runge-Kutta scheme is used in a fractional step method for time advancement.

Each particle is described by its center of mass, $\mathbf{x}(t)$, and velocity, $\mathbf{v}(t)$, which evolve according to,
\begin{equation}
\frac{d\mathbf{x}^{(i)}}{dt} = \mathbf{v}^{(i)}, \hspace{0.1\linewidth} \frac{d\mathbf{v}^{(i)}}{dt} = \mathbf{a}\left(\mathbf{x}^{(i)},t,\mathbf{v}^{(i)}\right).
\label{eq:particle-tracking}
\end{equation}
where $\mathbf{a}$ represents acceleration due to fluid forces. The Stokes drag force, $\mathbf{a} = \mathbf{a}_{St} = \tau_p^{-1} \left( \mathbf{u}(\mathbf{x}^{(i)}, t) - \mathbf{v}^{(i)} \right)$, applies when the particle Reynolds number is very small, $Re_p = |\mathbf{u} - \mathbf{v}| d_p / \nu_f \ll 1$. The particle adjusts to the background flow on the timescale, $\tau_p = \rho_p d_p^2 / (18 \mu_f)$. A better model at finite Reynolds number is the Schiller-Naumann drag \citep{Schiller1933, Balachandar2010}, $\mathbf{a} = \mathbf{a}_{SN} = \mathbf{a}_{St} \left( 1 + 0.15 Re_p^{0.687} \right)$, which is used in all the simulations in this paper. Particles are characterized by the Stokes number, $St^+ = \tau_p / \tau_*$, particle-to-fluid density ratio, $\rho_p / \rho_f$, and non-dimensional diameter, $d_p^+ = d_p / \delta_*$. Only two of these dimensionless parameters are independent,
\begin{equation}
St^+ = \frac{1}{18} \frac{\rho_p}{\rho_f} {d_p^+}^2.
\end{equation}

The particle phase is computed by advancing \eqref{eq:particle-tracking} with a second-order Runge-Kutta method, using trilinear interpolation to compute the fluid velocity at particle locations. Higher-order Lagrange interpolation schemes are tested and resulted in little change to the quantities of interest in this paper \citep{Johnson2019}. Particle-wall collisions are computed using a unity restitution coefficient. Particle-particle collisions were computed using a hard-sphere collision model with a specified restitution coefficient. Unless otherwise stated, particle-particle collisions also used a unity restitution coefficient. Particle and collision statistics shown in this paper are computed in wall-normal discretized bins of width $0.5$ viscous units, which equal to the minimum grid spacing of the Eulerian fluid solution.

\subsection{Particle statistics \label{sec:statistics}}
The (single-particle) statistics of monodisperse particles are described by the particle distribution function,
\begin{equation}
f\left(\mathbf{x}, \mathbf{v}; t\right) = \left\langle \sum_{i=1}^{N_p} \delta\left( \mathbf{x} - \mathbf{x}^{(i)}(t) \right) \delta\left( \mathbf{v} - \mathbf{v}^{(i)}(t) \right)  \right\rangle,
\label{eq:pdf-def}
\end{equation}
where $\delta(\mathbf{x})$ is the Dirac delta function and $N_p$ is the number of particles in the domain. In this work, $N_p$ remains constant in time, and the numerical simulations reintroduce particles exiting the domain at the complementary location along the opposite boundary, consistent with the periodic boundary conditions. The evolution of the particle statistics is governed by,
\begin{equation}
\frac{\partial f}{\partial t} + \nabla_{\mathbf{x}} \cdot \left( \mathbf{v} f \right) + \nabla_{\mathbf{v}} \cdot \left( \left\langle \left. \mathbf{a}\right| \mathbf{x}, \mathbf{v} \right\rangle f \right) = \dot{f}_{coll},
\label{eq:Boltzmann}
\end{equation}
where $\dot{f}_{coll} = \dot{f}_{coll}(\mathbf{x}, \mathbf{v}; t)$ represents the change in particle velocities caused by particle-particle collisions.

The particle concentration is defined by
\begin{equation}
C(\mathbf{x}; t) = \int f(\mathbf{x}, \mathbf{v}; t)  d\mathbf{v}.
\label{eq:concentration}
\end{equation}
The bulk concentration of particles is
\begin{equation}
C_0 = \frac{N_p}{V} = \frac{1}{V} \int C(\mathbf{x}; t) d\mathbf{x} =  \frac{1}{V} \iint f(\mathbf{x},\mathbf{v}; t) d\mathbf{v} d\mathbf{x},
\label{eq:bulk-concentration}
\end{equation}
where $V$ is the volume of the domain. The volume fraction and mass fraction characterizing the particle ensemble are, respectively,
\begin{equation}
\Phi_V = \tfrac{\pi}{6} d_p^3 C_0, \hspace{0.1\linewidth} \Phi_M = \tfrac{\rho_p}{\rho_f} \Phi_V.
\label{eq:bulk-fractions}
\end{equation}
Local volume and mass fractions can also be defined,
\begin{equation}
\phi_v =  \tfrac{\pi}{6} d_p^3 C, \hspace{0.1\linewidth} \phi_m = \tfrac{\rho_p}{\rho_f} \phi_v.
\label{eq:local-fractions}
\end{equation}

Integrating \eqref{eq:Boltzmann} in velocity and dividing by volume gives the evolution equation for the concentration field,
\begin{equation}
\frac{\partial C}{\partial t} + \nabla_{\mathbf{x}}\cdot\left( \langle \mathbf{v} | \mathbf{x} \rangle C \right) = 0.
\end{equation}
The collisional term vanishes because individual collisions do not add or remove particles (mass conserving). At steady state in the channel flow, this implies,
\begin{equation}
\frac{d\left( \langle v_y | y \rangle C\right)}{dy} = 0,
\end{equation}
which means that $\langle v_y | y \rangle = 0$.

\subsection{Turbophoresis \label{sec:turbophoresis}}
Likewise, an evolution equation expressing the conservation of momentum for the particle phase can be obtained by multiplying \eqref{eq:Boltzmann} by $\mathbf{v}$ before integrating in velocity,
\begin{equation}
\frac{\partial \left( \langle \mathbf{v} | \mathbf{x} \rangle C \right)}{\partial t} + \nabla_{\mathbf{x}} \cdot \left( \langle \mathbf{v} \otimes \mathbf{v} | \mathbf{x} \rangle C \right) - \langle \mathbf{a} | \mathbf{x} \rangle C = 0
\end{equation}
For the steady-state channel flow, this implies,
\begin{equation}
\frac{d\left( \langle v_y^2 | y \rangle C \right)}{dy} - \langle a_y | y \rangle C = 0,
\end{equation}
or
\begin{equation}
\langle v_y^2 | y \rangle \frac{dC}{dy} = \left( \langle a_y | y \rangle - \frac{d\langle v_y^2 | y \rangle}{dy} \right) C.
\label{eq:particle_momentum}
\end{equation}
The two right side terms represent two causes of non-uniform concentration profiles in the turbulent channel flow. The first term is the average wall-normal force on all particles at a particular distance from the wall. Using Stokes drag to illustrate the basic idea, $\langle a_y | y \rangle = \tau_p^{-1} \langle u_y | y \rangle$ (recall that $\langle v_y | y \rangle = 0$ at steady-state), where $\langle u_y | y \rangle > 0$ as inertial particles tend to preferentially sample ejection events \citep{Johnson2018,Johnson2019}. This biased sampling of the flow leads to a net force on particles away from the wall. The biased sampling force counteracts the turbophoresis pseudo-force, the second term on the right, which pushes particles toward the wall, down the gradient of particle wall-normal velocity variance, $\tfrac{d}{dy}\langle v_y^2 | y \rangle$. Equation \eqref{eq:particle_momentum} can be formally solved for the concentration,
\begin{multline}
C(y) = \mathcal{N} \exp\left( \int_{~}^{y} \frac{\langle a_y | \eta \rangle}{\langle v_y^2 | y \rangle} d\eta - \int_{~}^{y} \frac{d \langle v_y^2 | \eta \rangle}{d\eta} d\eta \right)\\
= \frac{\mathcal{N}}{\langle v_y^2 | \eta \rangle} \exp\left( \int_{~}^{y} \frac{\langle a_y | \eta \rangle}{\langle v_y^2 | y \rangle} d\eta \right),
\label{eq:concentration-momentum}
\end{multline}
where $\mathcal{N}$ is an integration constant for normalizing the integral of $C / C_0$. The concentration profile can thus be written in terms of phoresis integrals for biased sampling and turbophoresis,
\begin{equation}
I_{bias} = \int_{~}^{y} \frac{\langle a_y | \eta \rangle}{\langle v_y^2 | y \rangle} d\eta, \hspace{0.04\linewidth} I_{turb} = \int_{~}^{y} \frac{d \langle v_y^2 | \eta \rangle}{d\eta} d\eta.
\label{eq:phoresis-integrals}
\end{equation}
These phoresis integrals allow a direct evaluation of turbophoresis and biased sampling in determining the concentration profile.

\begin{figure}[t]
	\centering
	(a)\includegraphics[width=0.66\linewidth]{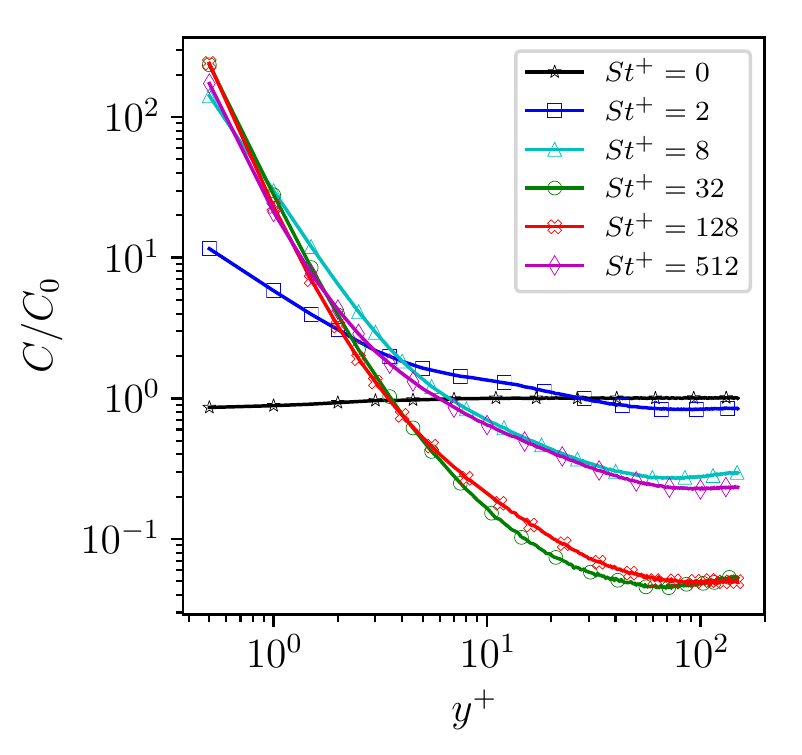}\\
	(b)\includegraphics[width=0.66\linewidth]{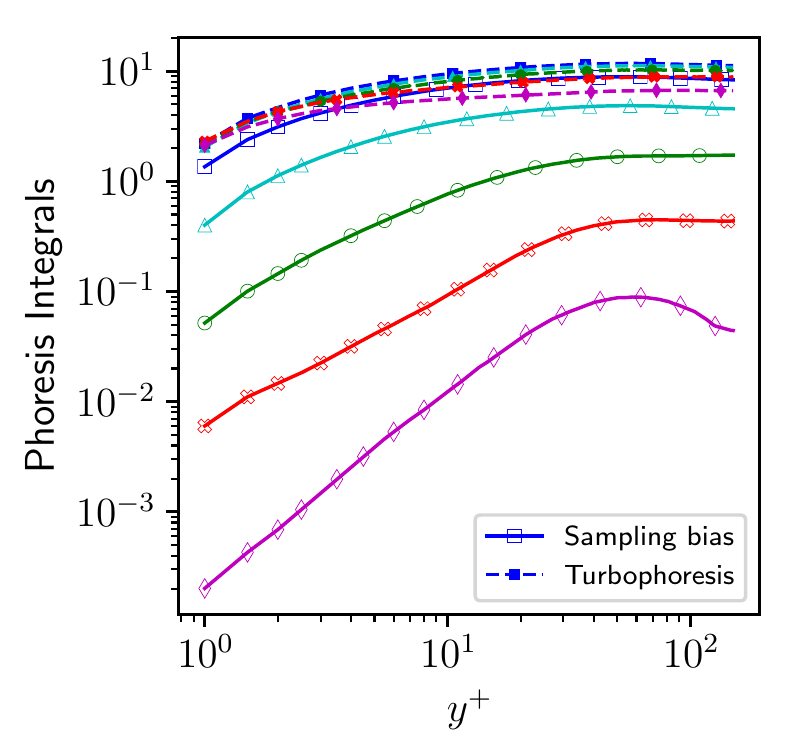}
	\caption{(a) Relative concentration and (b) phoresis integrals, Eq.\ \eqref{eq:phoresis-integrals}, of particles for various $St^+$ with $d_p^+ = 0.5$ and $\Phi_V = 0$ as a function of wall-normal distance in a $Re_* = 150$ channel flow.}
	\label{fig:Phi0}
\end{figure}

Figure \ref{fig:Phi0}(a) shows simulation results for the concentration profiles for a wide range of $St^+$ (holding $d_p^+ = 0.5$ constant and changing the density ratio). The simulations shown here neglect two-way coupling and particle-particle collisions, effectively representing the case of $\Phi_V = 0$. The $St^+ = 0$ particles (Lagrangian particles) have a uniform distribution, though a slight deviation in the viscous sublayer illustrates the minor effects of discretization and interpolation errors. At finite but small $St^+ = 2$, turbophoresis begins to raise the concentration in the near-wall region up to $10$ times the bulk concentration. As $St^+$ is further increased, the near-wall concentration exceeds $100$ times the mean concentration.

The change in concentration profile with $St^+$ is analyzed in Figure \ref{fig:Phi0}(b) using phoresis integrals, Eq. \eqref{eq:phoresis-integrals}. At $St^+ = 2$, the biased sampling force is comparable in magnitude to turbophoresis, attenuating the large near-wall concentrations seen at higher $St^+$. In fact, it can be shown analytically that the biased sampling term exactly cancels turbophoresis in the $St^+ \rightarrow 0$ limit \citep{Johnson2018,Johnson2019}. As $St^+$ increases, both turbophoresis and biased sampling decrease in magnitude, though turbophoresis decreases much more slowly. The result is the predominance of turbophoresis over biased sampling at higher $St^+$, leading to $C(y) \sim 1/\langle v_y^2 | y \rangle$, Eq. \eqref{eq:concentration-momentum}.

\subsection{Particle-particle collisions \label{sec:collisions}}

\begin{figure*}[t]
	\centering
	\includegraphics[width=0.32\linewidth]{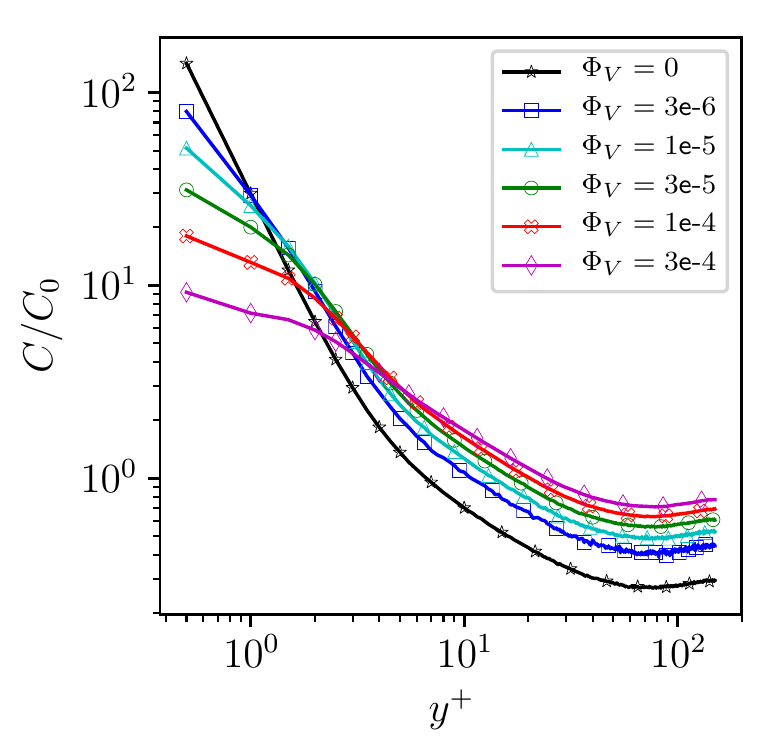}
	\includegraphics[width=0.33\linewidth]{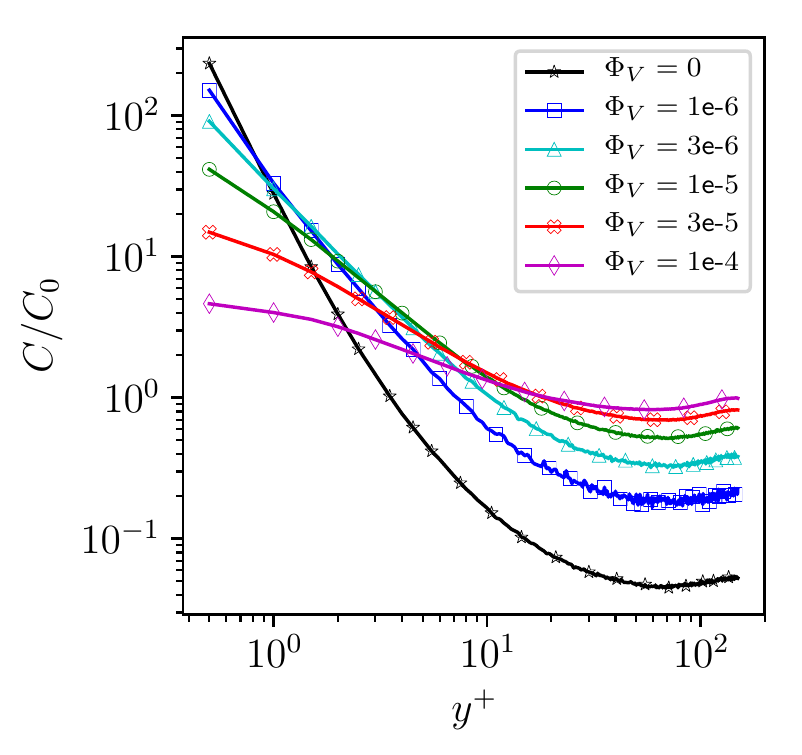}
	\includegraphics[width=0.33\linewidth]{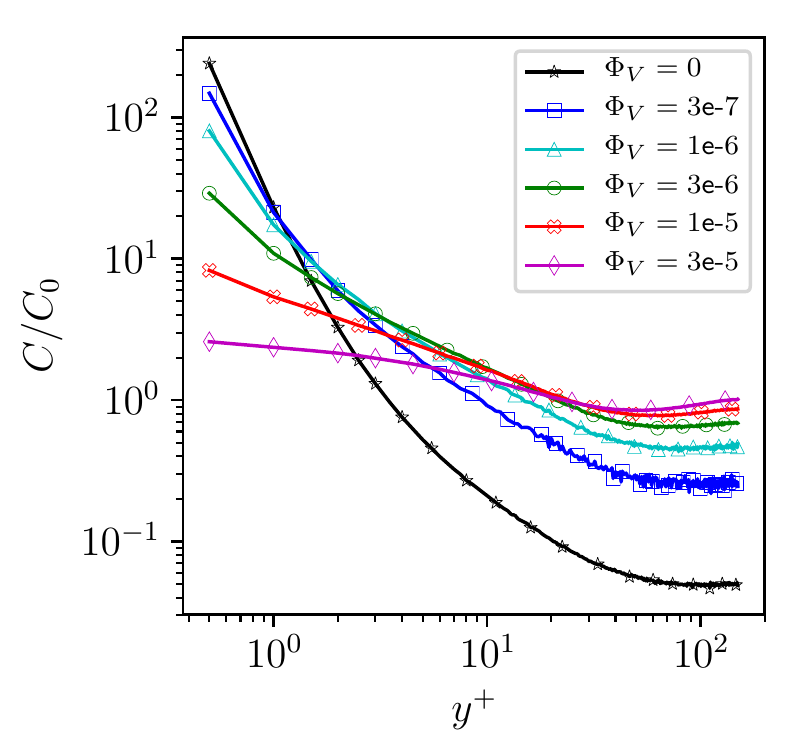}
	%~\includegraphics[width=0.32\linewidth]{C_avg_Re150_St512.pdf}\\
	(a) \hspace{0.325\linewidth} (b) \hspace{0.325\linewidth} (c)
	\caption{Particle concentration profiles for various volume fractions, $\Phi_V$, at (a) $St^+ = 8$, (b) $St^+ = 32$, (c) $St^+ = 128$.}
	\label{fig:wcollSt}
\end{figure*}

\begin{figure*}[t]
	\centering
	\includegraphics[width=0.325\linewidth]{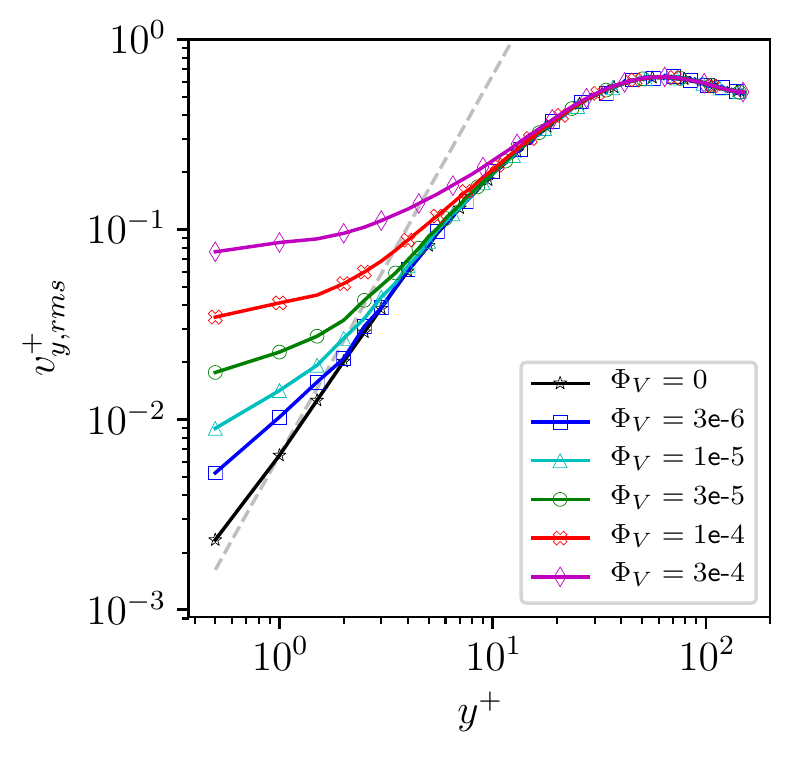}
	\includegraphics[width=0.325\linewidth]{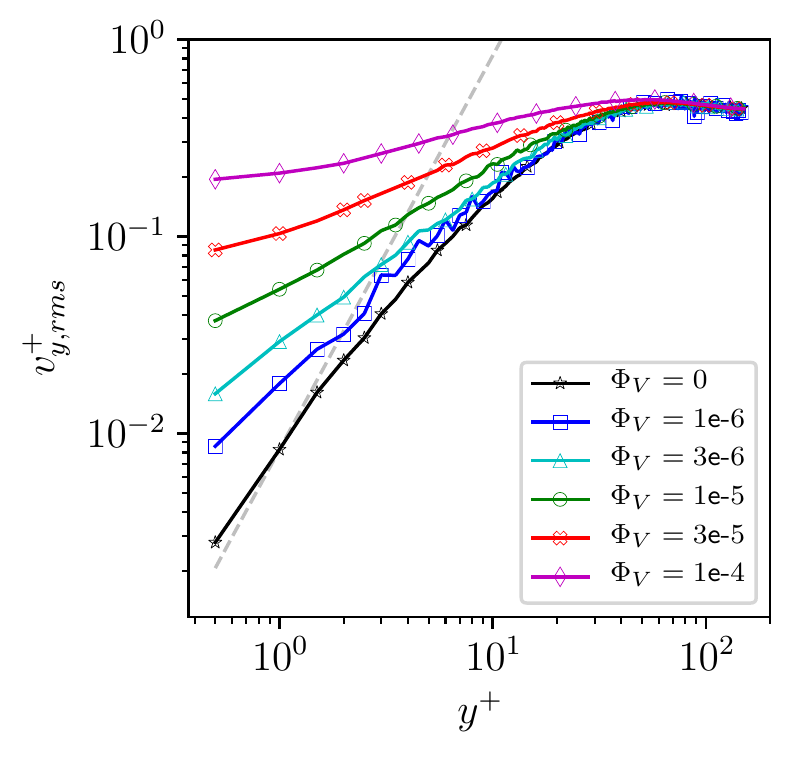}
	\includegraphics[width=0.325\linewidth]{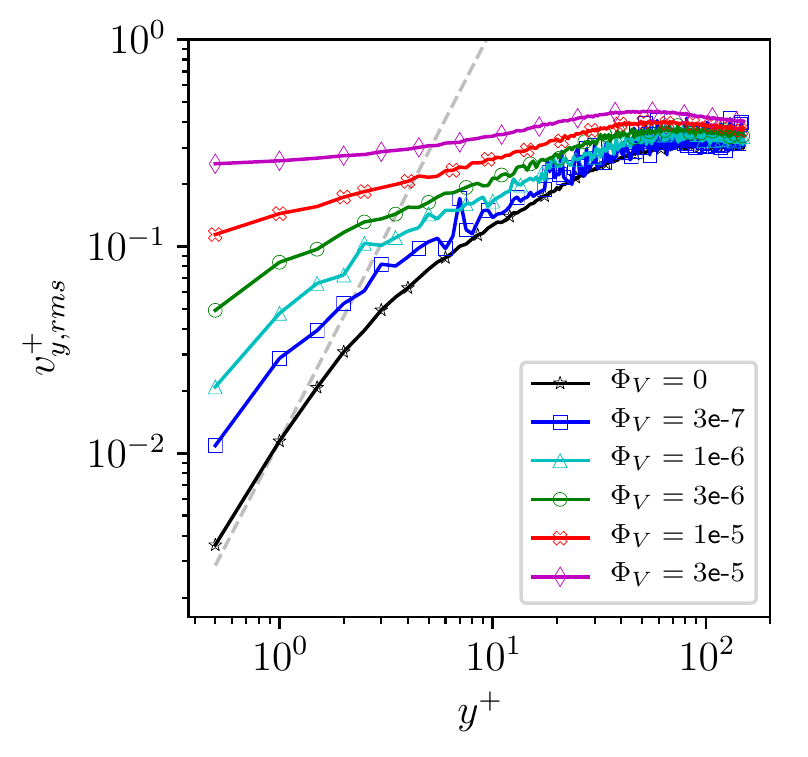}\\
	(a) \hspace{0.325\linewidth} (b) \hspace{0.325\linewidth} (c)
	\caption{Particle wall-normal velocity root-mean-square profiles for various volume fractions, $\Phi_V$, at (a) $St^+ = 8$, (b) $St^+ = 32$, (c) $St^+ = 128$. The gray dashed lines indicate the $v \sim y^2$ asymptotic behavior for fluid particles near the wall.}
	\label{fig:wcollSt-vy}
\end{figure*}

Particle-particle collisions are known to effect particle dynamics in wall-bounded turbulent flows as volume fractions become high enough \citep{Li2001, Yamamoto2001, Caraman2003, Kuerten2015, Kuerten2016a, Johnson2019}. In fact, the volume fraction at which collisions have a significant effect on the concentration profile can be quite `low', as shown in Figure \ref{fig:wcollSt}. Relative concentration profiles for three different $8 \leq St^+ \leq 128$ are shown in the four panels, each one containing a range of $\Phi_V$. As the $\Phi_V$ is increased, the near-wall peak in concentration is attenuated. At high enough $\Phi_V$, the concentration profile becomes almost flat. The volume fractions shown for each $St^+$ value in Figure \ref{fig:wcollSt} and \ref{fig:wcollSt-vy} were chosen to best capture the transition away from the $\Phi_V = 0$ concentration profile.

These changes in the relative concentration profiles with $\Phi_V$ reflect how particle-particle collisions alter turbophoresis. Indeed, the main effect of increasing $\Phi_V$ is to increase the rate at which a given particle collides with another particle. Each collision redistributes the participating particles in velocity-space, impacting $\langle v_y^2 | y \rangle$ and hence changing how the turbophoresis forms the concentration profile. On average, particle-particle collisions near the wall transfer particle fluctuation energy from the streamwise to wall-normal components (return-to-isotropy for particle velocity fluctuations), enhancing the near-wall $\langle v_y^2 | y \rangle$ and decreasing the turbophoretic drift \citep{Caraman2003, Johnson2019}. The resulting profiles of $\langle v_y^2 | y \rangle$ are shown in Figure \ref{fig:wcollSt-vy} for the same $St^+$ and $\Phi_V$ in Figure \ref{fig:wcollSt}. At low $\Phi_V$, the particle velocities near the wall follow closely the asymptotic behavior of fluid velocities, i.e., $v_y \sim y^2$. This is particularly true at lower $St^+$. As $\Phi_V$ increases, the near-wall $\langle v_y^2 | y \rangle$ increases dramatically due to particle-particle collisions, resulting in the reduction of the near-wall concentration peak seen in Figure \ref{fig:wcollSt}.

\begin{figure*}
	\centering
	\includegraphics[width=0.325\linewidth]{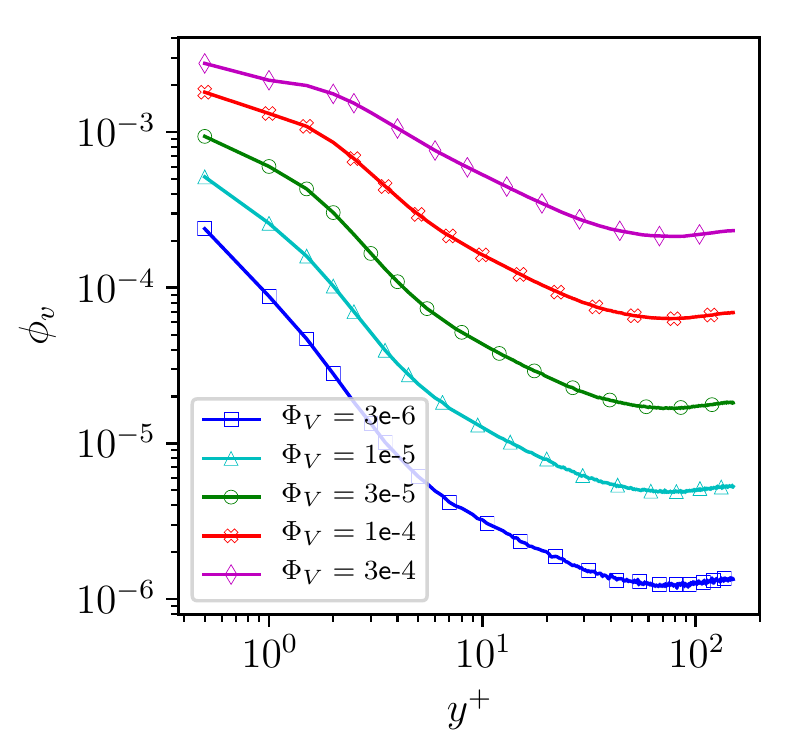}
	\includegraphics[width=0.325\linewidth]{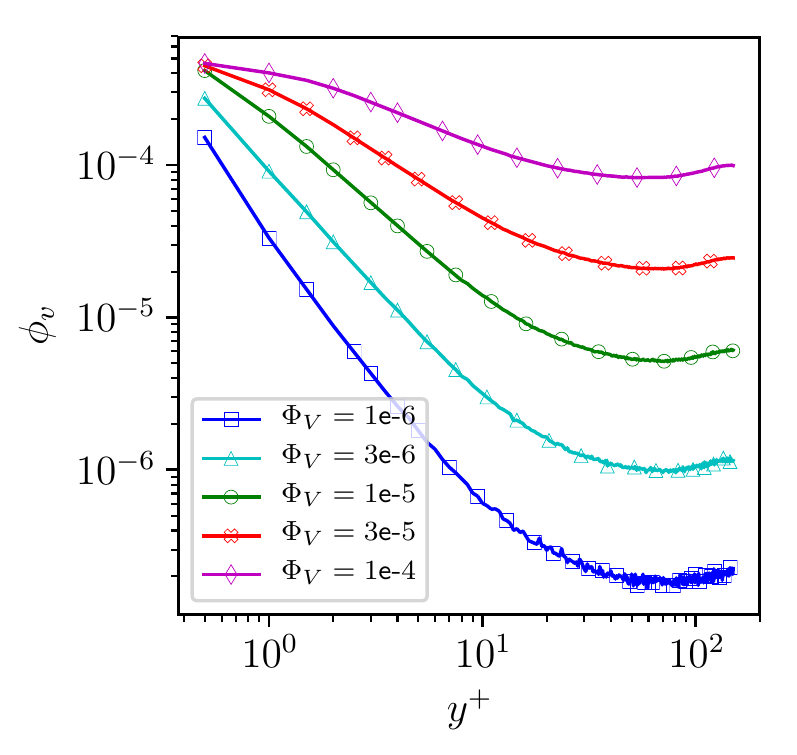}
	\includegraphics[width=0.325\linewidth]{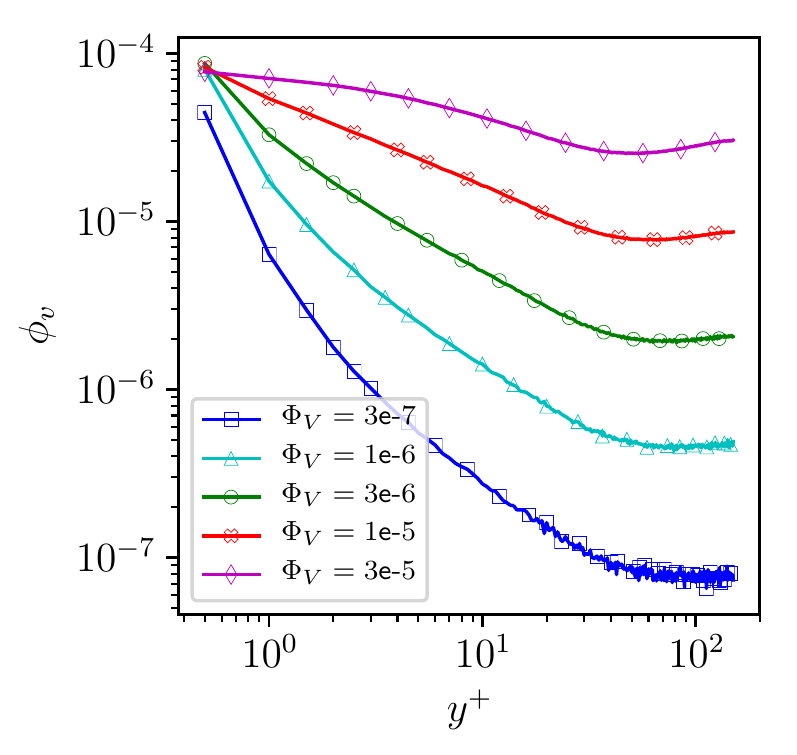}\\
	(a) \hspace{0.325\linewidth} (b) \hspace{0.325\linewidth} (c)
	\caption{Local volume fraction ($\phi_v$) profiles for various bulk volume fractions, $\Phi_V$, at (a) $St^+ = 8$, (b) $St^+ = 32$, (c) $St^+ = 128$.}
	\label{fig:wcollSt-PhiV}
\end{figure*}

It can be seen from Figures \ref{fig:wcollSt} and \ref{fig:wcollSt-vy} that collisions alter the relative concentration profile at quite low $\Phi_V$. While the trend with $\Phi_V$ is clearly demonstrated, a more detailed view of the relative importance of collisions can consider the local volume fraction, $\phi_v$. Figure \ref{fig:wcollSt-PhiV} plots $\phi_v$ as function of wall-normal distance. For $St^+ \geq 32$, the near-wall peak volume fraction appears to saturate even as the bulk volume fraction, $\Phi_V$, increases. This likely also occurs for lower $St^+$, but the simulations for this paper did not reach high enough $\Phi_V$ to observe it. As the volume (and mass) fraction continues to increase, at some point two-way coupling could alter this observation, but the two-way coupling has a negligible impact on the concentration profile for the simulation parameters shown in this paper \citep{Johnson2019}. The saturation of local volume fraction near the wall suggests that particle-particle collisions limit turbophoresis in such a way as to effectively set a maximum local volume fraction above which turbophoresis cannot sustain the near-wall peak. This maximum local volume fraction, $\phi_{v,max}$, evidently decreases with increasing $St^+$. A simple explanation for this trend is as follows.

Roughly speaking, $\phi_{v,max}$ may be related to the volume fraction as which collisional effects begin to dominate over drag in determining the trajectories and statistics of particles, e.g., $\langle v_y^2 | y \rangle$. The relative impact of collisions and drag can be quantified by the dimensionless parameter, $c = \tau_p / \tau_{coll}$, where $\tau_{coll}$ is the typical time between collisions for a given particle. If $c$ is small, particles have plenty of time in between collisions to adjust back to the velocity of the nearby fluid. If $c$ is large, however, particles behave ballistically between collisions and fluid drag play a minimal role in establishing particle statistics such as $\langle v_y^2 | y \rangle$. It follows, then, that the maximum (local) volume fraction that turbophoresis (alone) can support can be approximated by a critical value of this parameter, $c = c_{cr} \sim \mathcal{O}(1)$. For wall-bounded flows, $\tau_{coll} \sim \lambda_{mf} / u_*$, where $\lambda_{mf}$ is the mean-free-path of the particles and the friction velocity, $u_*$, establishes the typical relative velocity between a particle that has just collided and other nearby particles. If the swept volume between collisions, $\sim d_p^2 \lambda_{mf}$, is set equal to the typical volume containing one particle on average, $\sim d_p^3 / \phi_v$, then the mean-free-path may be estimated by $\lambda_{mf} \sim d_p / \phi_v$. Combining these approximations, $c \sim St^+ \phi_v / d_p^+ \sim d_p^+ \phi_m$. In this way, it can be argued that the maximum volume fraction for significant collisional effects scales as $\phi_{v,max} \sim d_p^+ / St^+$. This scaling is roughly observed in Figure \ref{fig:wcollSt}, though certainly a more nuanced picture could be gained from more detailed analysis which is not pursued here. As a result of this trend, that $\phi_{v,max}$ decreases with $St^+$, collisional effects become noticeable in the relative concentration profiles at lower $\Phi_V$ as $St^+$ increases. As a result, the deviation from $\Phi_V = 0$ behavior in Figures \ref{fig:wcollSt} and \ref{fig:wcollSt-vy} begins at lower volume fractions as $St^+$ increases. This qualitative trend was recognized by \citet{Elghobashi1994} when updating the regime diagram from \citet{Elghobashi1991} to include four-way coupling at lower volume fractions as the particle Stokes number increases.

\section{Reducing the number of computational particles: `super-particles'\label{sec:theory}}
In the previous section, it was shown how turbophoresis generates a near-wall peak in particle concentration. In particular, the impact of particle-particle collisions on this process was explored, showing that the relative concentration profile in a wall-bounded turbulent flow can depend strongly on the volume fraction. Therefore, reducing the number of computational particles in the simulation will lead to unphysical changes to the particle concentration field if the collision treatment is not modified. This section introduces a simple method for recovering collisional effects without resorting to stochastic collision models.

The statistical framework of \S \ref{sec:statistics} is modified in \S \ref{sec:super-statistics} to address the situation where a reduced number of computational particles are simulated, with each computational particle carrying a statistical weight signifying that it effectively represents multiple physical particles. Here, these computational particles are referred to as `super-particles'. In other contexts, they are called particle `parcels' \citep{Subramaniam2013}. In \S \ref{sec:super-collisions}, collisions between super-particles are considered, leading to proposed scaling laws for enhancing the collision radius of super-particles in \S \ref{sec:super-size} in the limit of low and high Stokes numbers.

\subsection{Super-particle statistics \label{sec:super-statistics}}
A super-particle denotes a computational particle which does not necessarily represent one physical particle, but represents more than one physical particle through the concept of a statistical weight assigned to each particle. In particle-laden flow simulations with Lagrangian tracking, Eq.\ \eqref{eq:particle-tracking}, the computational cost of simulating the particle phase scales with $N_p$, the number of physical particles in the system of interest. The super-particle approach instead tracks $N_c$ computational particles, whose statistics are meant to be representative of the full physical system by assigning each super-particle a statistical weight of $W^{(i)}$.  In some sense, the super-particle approach for the particle phase is akin to large-eddy simulations (LES) for the fluid phase, in that a description of the system with a reduced number of computational degrees of freedom is sought.

The statistics of the super-particles are described by,
\begin{equation}
g\left(\mathbf{x}, \mathbf{v}; t\right) = \left\langle \sum_{i=1}^{N_c} W^{(i)} \delta\left( \mathbf{x} - \mathbf{x}^{(i)}(t) \right) \delta\left( \mathbf{v} - \mathbf{v}^{(i)}(t) \right)  \right\rangle,
\label{eq:super-pdf-def}
\end{equation}
which reduces to \eqref{eq:pdf-def} when $W^{(i)} = 1$ and $N_c = N_p$. The present work follows and extends the theoretical approach of \citet{Garg2009} and \citet{Subramaniam2013}, which used `$h$' to denote this super-particle distribution function -- a notation which is not followed here for reasons that will become obvious later. For now, the statistical weight attached to each particle, $W^{(i)}$, will be kept constant. Further extension along the lines of \citet{Garg2009} to vary $W^{(i)}$ in order to achieve faster numerical convergence is possible but not pursued at the present. All single-particle statistics can be written in terms of $g$. For example, the concentration is,
\begin{equation}
C(\mathbf{x}; t) = \frac{1}{V} \int g(\mathbf{x}, \mathbf{v}; t)  d\mathbf{v},
\end{equation}
and
\begin{equation}
C_0(t) = \frac{N_p}{V} = \int C(\mathbf{x}; t) d\mathbf{x} = \frac{N_c W}{V},
\end{equation}
which fixes the correct choice of $W = N_p / N_c$. Thus, $W$ represents the factor by which the number of computational particles is reduced. The volume and mass fractions can be defined using Eqs. \eqref{eq:bulk-fractions} and \eqref{eq:local-fractions}.

The trajectory of each individual super-particle is described by Eq. \eqref{eq:particle-tracking}. The evolution of the super-particle distribution function is,
\begin{equation}
\frac{\partial g}{\partial t} + \nabla_{\mathbf{x}} \cdot \left( \mathbf{v} g \right) + \nabla_{\mathbf{v}} \cdot \left( \left\langle \left. \mathbf{a}\right| \mathbf{x}, \mathbf{v} \right\rangle g \right) = \dot{g}_{coll}.
\label{eq:super-Boltzmann}
\end{equation}
In order to achieve accurate statistics from the super-particle approach, the distribution functions $f$ and $g$ should have the same evolution for the same initial and boundary conditions. This is the principle of statistical equivalence \citep{Subramaniam2013}. Another way to state statistical equivalence is to say that the evolution of $g$ should be invariant with respect to changes in $W$. It is immediately clear, then, that the drag law used for super-particles, $\mathbf{a}$, must be kept invariant with respect to $W$. That is, the super-particles must use the same drag law as physical particles.

It is evident from Figures \ref{fig:wcollSt} and \ref{fig:wcollSt-vy}, along with the discussion of the underlying physics, that the particle-particle collisions represented by $\dot{g}_{coll}$ break the invariance of \eqref{eq:super-Boltzmann} with respect to $W$. That is, if a reduced number of computational particles are simulated with the same drag law and same collision model, the result is as if a lower volume fraction is simulated. In this way, the variation of relative concentration, $C(y) / C_0$, and particle wall-normal velocity variance, $\langle v_y^2 | y \rangle$, shown in Figures \ref{fig:wcollSt} and \ref{fig:wcollSt-vy} respectively may be interpreted as errors of a naive super-particle approach. A closer look at $\dot{g}_{coll}$ and how its breaks invariance with respect to $W$ is warranted.

\subsection{Super-particle collisions \label{sec:super-collisions}}

\begin{figure}
	\centering
	\includegraphics[trim = 0mm 62mm 0mm 0mm, clip, width=1.0\linewidth]{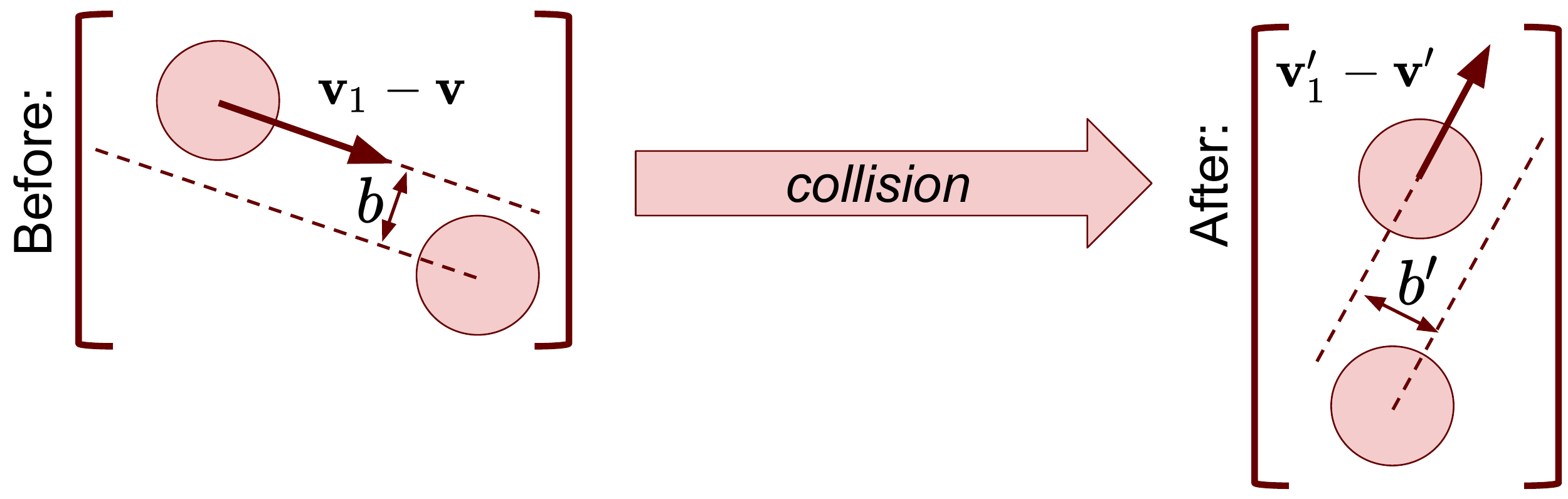}
	\caption{Simplified illustration of a representative binary collision.}
	\label{fig:collision-schematic}
\end{figure}

The analysis pursued below treats only binary collisions. Collisions involving three or more particles are assumed to be rare enough so as to be negligible in their impact on the quantities of interest. A sketch of a representative binary collision is shown in Figure \ref{fig:collision-schematic}. A reference particle with pre-collision velocity $\mathbf{v}$ is impacted by a colliding particle with pre-collision velocity $\mathbf{v}_{1}$. The particles collide with an offset $b < d_p$, which determines the obliqueness of the collision. After the collision, the particles have velocities $\mathbf{v}^{\prime}$ and $\mathbf{v}_{1}^{\prime}$ respectively. The post-collision velocities depend on the restitution of coefficient. The collisional source term in Eq. \eqref{eq:Boltzmann} may be modeled as,
\begin{equation}
\dot{f}_{coll} = \iint_{0}^{d_p} b db d\theta \int d\mathbf{v}_{1} |\mathbf{v}_1 - \mathbf{v}| \left[ f(\mathbf{v}^\prime) f(\mathbf{v}_1^\prime)  - f(\mathbf{v}) f(\mathbf{v}_1) \right].
\label{eq:collision-source}
\end{equation}
This form is a bit simplified, for example, in assuming that the velocities of the colliding particles are statistically independent. Removing this assumption requires knowledge of two-particle statistics, a richer description than provided by Eq. \eqref{eq:pdf-def} or \eqref{eq:super-pdf-def}. Nonetheless, Eq. \eqref{eq:collision-source} illustrates the basic scaling of $\dot{f}_{coll}$, which is all that is used for the modeling in this paper. In particular,
\begin{equation}
\dot{f}_{coll} \sim \delta v_p ~d_p^2 ~N_p^2,
\label{eq:collision-scaling}
\end{equation}
where $\delta v_p$ is the the relative velocity at collision, i.e., relative velocity when the particles center of masses are separated by a distance $d_p$. More generally, it can be seen that $\dot{f}_{coll}$ scales with the collision rate, i.e., the number of collisions per unit time. Indeed, previous estimates of the collision rate \citep{Sundaram1997,Wang2000} lead to Eq. \eqref{eq:collision-scaling} as well, though may differ from each other in whether the relative velocity magnitude or relative radial velocity is used for $\delta v_p$.

In the same way, the collisional source term in Eq. \eqref{eq:super-Boltzmann} may be modeled as,
\begin{equation}
\dot{g}_{coll} = \iint_{0}^{d_c} b db d\theta \int d\mathbf{v}_{1} |\mathbf{v}_1 - \mathbf{v}| \left[g(\mathbf{v}^\prime) \frac{g(\mathbf{v}_1^\prime)}{W}  - g(\mathbf{v}) \frac{g(\mathbf{v}_1)}{W} \right],
\label{eq:super-collision-source}
\end{equation}
or
\begin{equation}
\dot{g}_{coll} \sim \delta v_c d_c^2 N_p^2 / W.
\label{eq:super-collision-scaling}
\end{equation}
In other words, if the collisional diameter of the super-particles, $d_c$, is kept equal to the physical diameter of the particles, $d_p$, and the relative velocity statistics are assumed unchanged, the collision rate will be under-estimated proportional to $W^{-1}$. The underlying reason is as follows. The single-particle distribution, Eq.\ \eqref{eq:pdf-def}, scales linearly with the number of particles, so that particle statistics can be represented by a reduced number of particles using the (linear) correction factor $W = N_p / N_c$. The binary collision rate, and hence $\dot{f}_{coll}$, scales quadratically with the number of particles. Therefore, the linear correction factor $W$ in Eq.\ \eqref{eq:super-pdf-def} only corrects for one of these two factors.

Said another way, each super-particle statistically represents $W$ physical particles, and hence when it does collide with another particle, this effectively represents the effect of $W$ particle-particle collisions. However, all other features held constant (relative velocity statistics, preferential concentration, etc...), the collision rate falls by a factor of $W^{-2}$ in the super-particle system. Therefore, the net effect of collisions on the resulting simulation statistics falls by a factor of $W^{-1}$, Eq. \eqref{eq:super-collision-scaling}. Without further treatment, the effect of collisions will not be represented accurately in the super-particle approach leading to errors in, for instance, concentration (Figure \ref{fig:wcollSt}) and velocity (Figure \ref{fig:wcollSt-vy}).

\subsection{Enhanced collision radius model\label{sec:super-size}}

A simple approach to modeling collisional effects in a super-particle approach is to try to restore invariance of $\dot{g}_{coll}$ with respect to $W$. One way this can be done is by enhancing the collisional diameter of the particles to be larger than the physical diameter of the particles, $d_c > d_p$. Because the velocity at collision, $\delta v_c$, may depend on the collisional diameter, a H\"{o}lder exponent-like relation is assumed for the velocity difference between two particles separated by a distance $d_c$,
\begin{equation}
v_c \sim d_c^h.
\label{eq:super-Holder}
\end{equation}
In the limit of very low Stokes number, the perturbation solution of \citet{Maxey1987} may be used to approximate the relative velocity of the particles in the vicinity of collision. Here the effects of two-way coupling, particle screening, and lubrication are not explicitly considered. In this approximation, the particle velocities are represented as a smooth, single-valued field, so that $h = 1$ in Eq. \eqref{eq:super-Holder}. In the opposite limit of high Stokes number, the colliding particles do not adjust at all to the local fluid velocity field as they approach. As a result, the collision velocity is independent of their separation at collision and $h = 1$ for \eqref{eq:super-Holder}. These two limits may be summarized as,
\begin{equation}
\lim\limits_{\tau_p \rightarrow 0} h = 1, \hspace{0.1\linewidth} \lim\limits_{\tau_p \rightarrow \infty} h = 0.
\label{eq:super-Holder-limits}
\end{equation}

With this approximation for the collision velocity, the super-particle collisional source term scales as,
\begin{equation}
\dot{g}_{coll} \sim d_c^{2+h} N_p^2 / W,
\end{equation}
and therefore the invariance of $\dot{g}_{coll}$ with respect to $W$ may be upheld by the choice,
\begin{equation}
d_c = d_p W^{1/(2+h)},
\label{eq:super-diac}
\end{equation}
which in the two limits discussed above results in,
\begin{equation}
\lim\limits_{\tau_p \rightarrow 0} \frac{d_c}{d_p} = \sqrt[3]{W}, \hspace{0.1\linewidth} \lim\limits_{\tau_p \rightarrow \infty} \frac{d_c}{d_p} = \sqrt{W}.
\end{equation}

As an example, consider a super-particle simulation with $64$ times fewer particles than the physical system under investigation. If the particles are at very low Stokes numbers, then the analysis here suggests scaling the collisional diameter by a factor of $4$ and proceeding with a deterministic collision algorithm. For very high Stokes number particles, this analysis suggests that more accurate results could be obtained by scaling the collision diameter by a factor of $8$.

The scaling for the collisional diameter, Eq. \eqref{eq:super-diac} along with Eq. \eqref{eq:super-Holder} implies a scaling for the collisional velocity,
\begin{equation}
\delta v_c = \delta v_p W^{h / (2+h)}.
\end{equation}
Because the change of momentum of particles due to collision is proportional to the relative (radial) velocity at collision, this ratio can be factored into the coefficient of restitution used for the collision of super-particles,
\begin{equation}
e_c = e_p W^{-h / (2+h)},
\label{eq:super-ec}
\end{equation}
where $e_p$ is the physical coefficient of restitution. The proposed model, then, consists of computing collisions based on an enhance collisional diameter according to Eq. \eqref{eq:super-diac} with a modified restitution coefficient from Eq. \eqref{eq:super-ec}. The model has a single parameter $h$ which must be chosen based on the expected behavior of the relative velocities of collisions (smooth vs. ballistic). While the collision radius is increased for particle-particle collisions, it is important to note that the treatment of particle-wall collisions continues to use the physical diameter of the particle, $d_p$, because particle-wall collision rates scale linearly with $N_c$ and naturally do not break invariance with $W$.

%\subsection{Collisional scaling analysis}
%scaling analysis in different regimes (ballistic vs low-St), when each scaling should apply based on part I

\section{Results: super-particles and turbophoresis \label{sec:results}}
This section demonstrates the performance of the proposed enhanced collision radius approach. In particular, the low and high Stokes number limits of Eq.\ \eqref{eq:super-Holder-limits} are verified in \S \ref{sec:St-limits}. Following that, a hybrid model bridging the two limits is introduced and validated in \S \ref{sec:hybrid-model}. The advantage of the hybrid model is that it computes a unique value of $h$ for each binary collision based on the known properties of the flow and particles participating. Therefore, the hybrid model performs well for low, intermediate, and high Stokes numbers. As motivated in the introduction and \S \ref{sec:background}, the concentration of particles as a function of wall-normal distance is treated as the main quantity of interest, though other features are shown when of particular interest.

\subsection{Limiting behaviors \label{sec:St-limits}}
The enhanced collision radius model in \S \ref{sec:super-size} can be directly employed by assuming a value for $h$. This direct application makes sense for $h = 1$ (for low $St^+$) or $h = 0$ (for high $St^+$) by essentially assuming that all particle-particle collisions in a flow occurs either at low or high inertia relative to the surrounding flow. As a first demonstration, the direct application of $h = 0$ and $h = 1$ is used for monodisperse particle-laden channel flows with three different values of $St^+$.

\begin{figure*}[t]
	\includegraphics[width=0.33\linewidth]{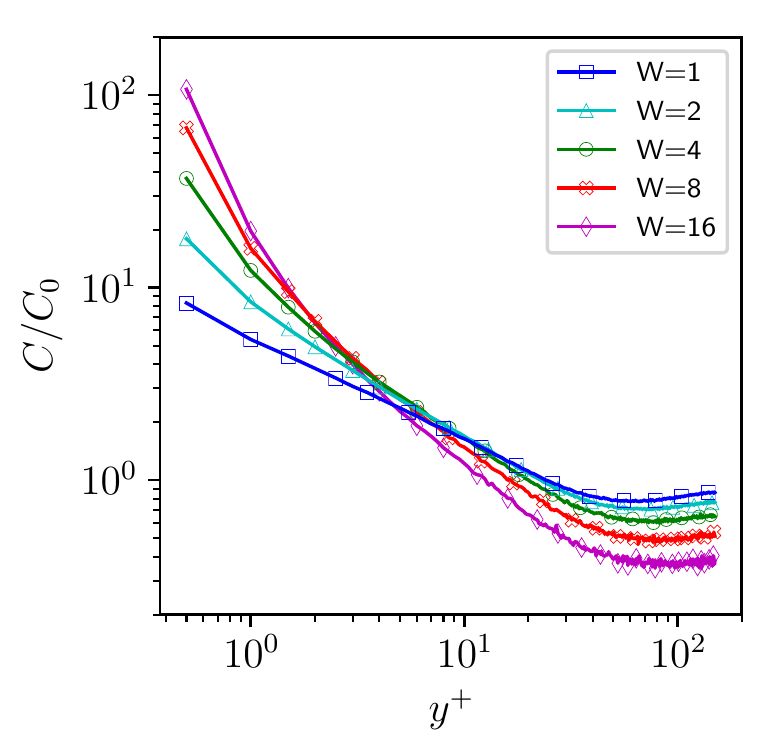}
	\includegraphics[width=0.33\linewidth]{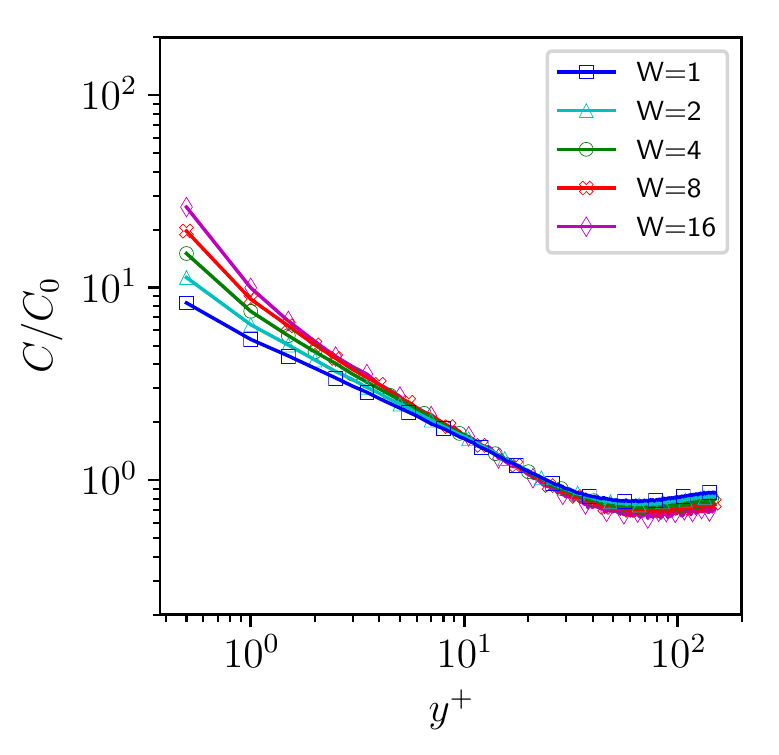}
	\includegraphics[width=0.33\linewidth]{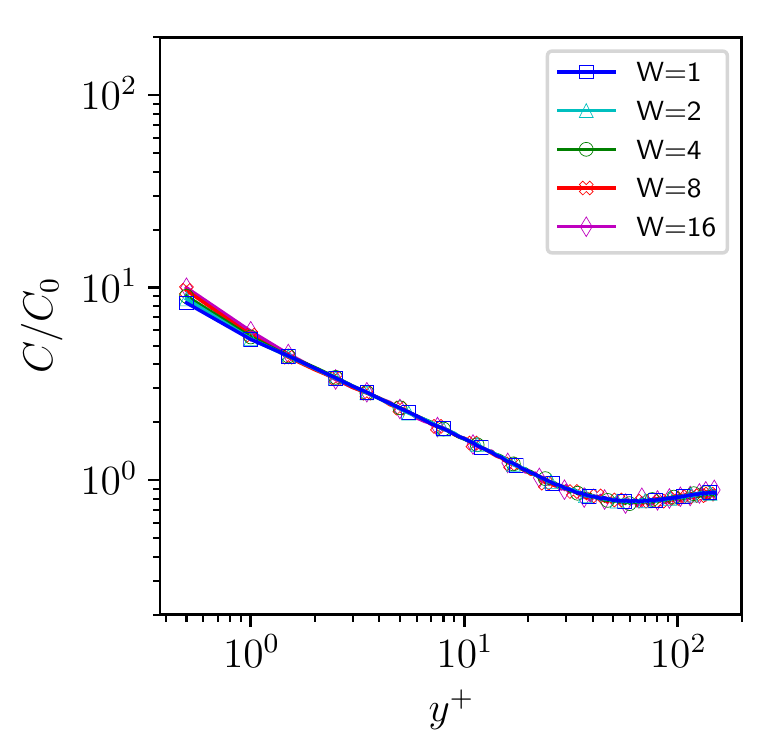}\\
	(a) \hspace{0.33\linewidth} (b) \hspace{0.33\linewidth} (c)
	\caption{Relative concentration profiles of $St^+ = 128$ particles at $\Phi_V = 1 \times 10^{-5}$, simulated with $N_c = N_p / W$ computational particles and (a) no correction, $d_c = d_p$; (b) low Stokes number correction, $h = 1$; and (c) high Stokes number correction, $h = 0$.}
	\label{fig:St128-h0}
\end{figure*}

\begin{figure*}[t]
	\includegraphics[width=0.33\linewidth]{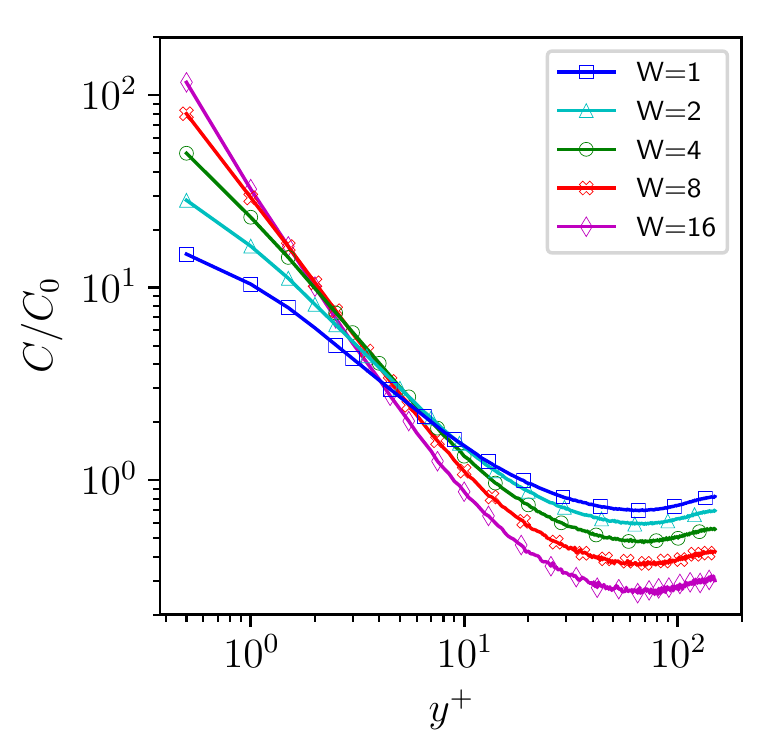}
	\includegraphics[width=0.33\linewidth]{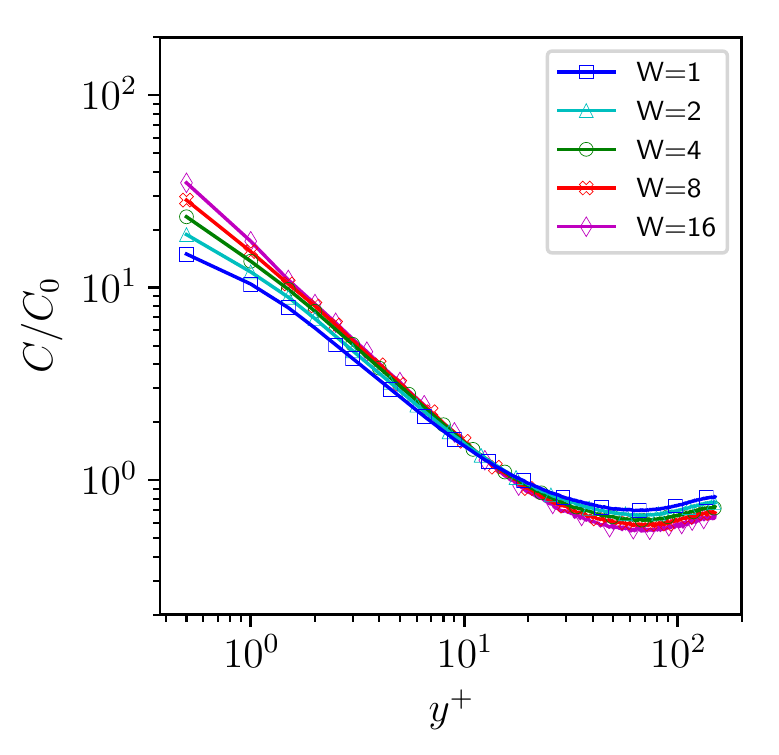}
	\includegraphics[width=0.33\linewidth]{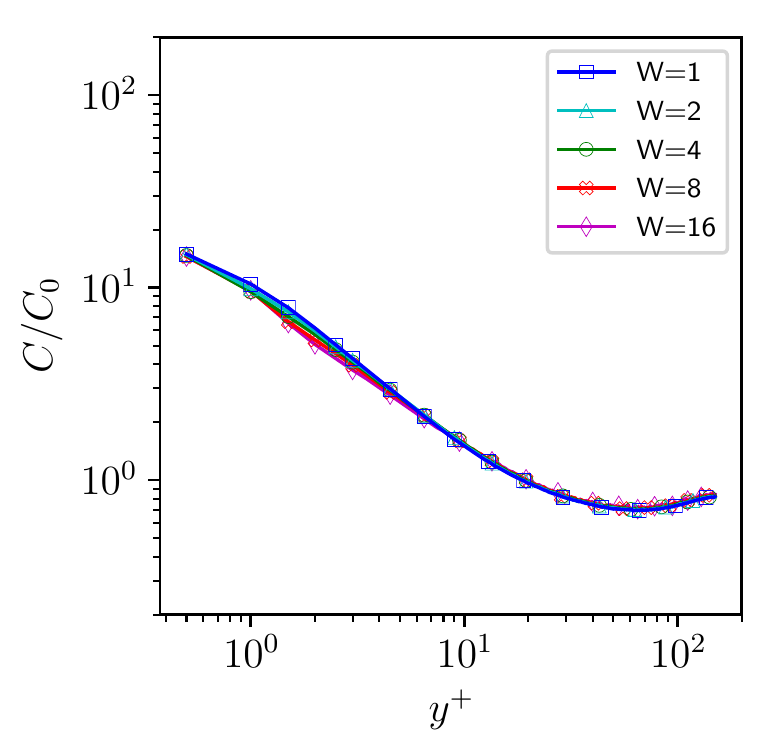}\\
	(a) \hspace{0.33\linewidth} (b) \hspace{0.33\linewidth} (c)
	\caption{Relative concentration profiles of $St^+ = 32$ particles at $\Phi_V = 3 \times 10^{-5}$, simulated with $N_c = N_p / W$ computational particles and (a) no correction, $d_c = d_p$; (b) low Stokes number correction, $h = 1$; and (c) high Stokes number correction, $h = 0$.}
	\label{fig:St32-h0}
\end{figure*}

\begin{figure*}[t]
	\includegraphics[width=0.33\linewidth]{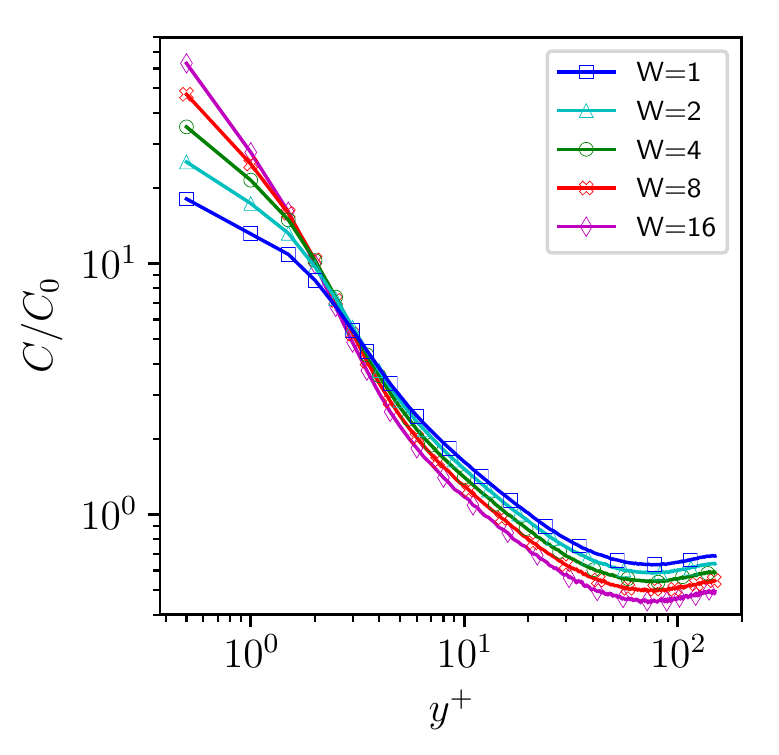}
	\includegraphics[width=0.33\linewidth]{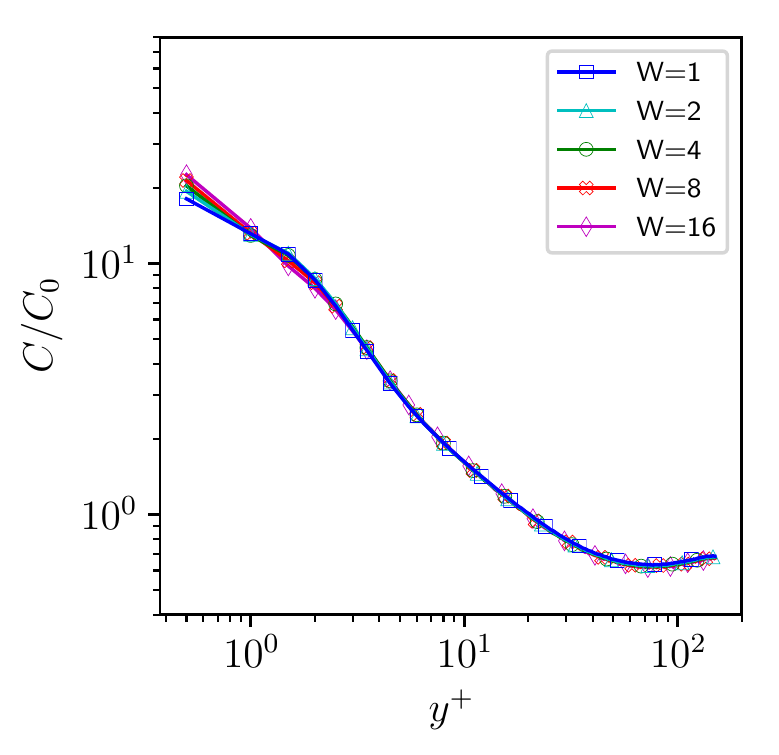}
	\includegraphics[width=0.33\linewidth]{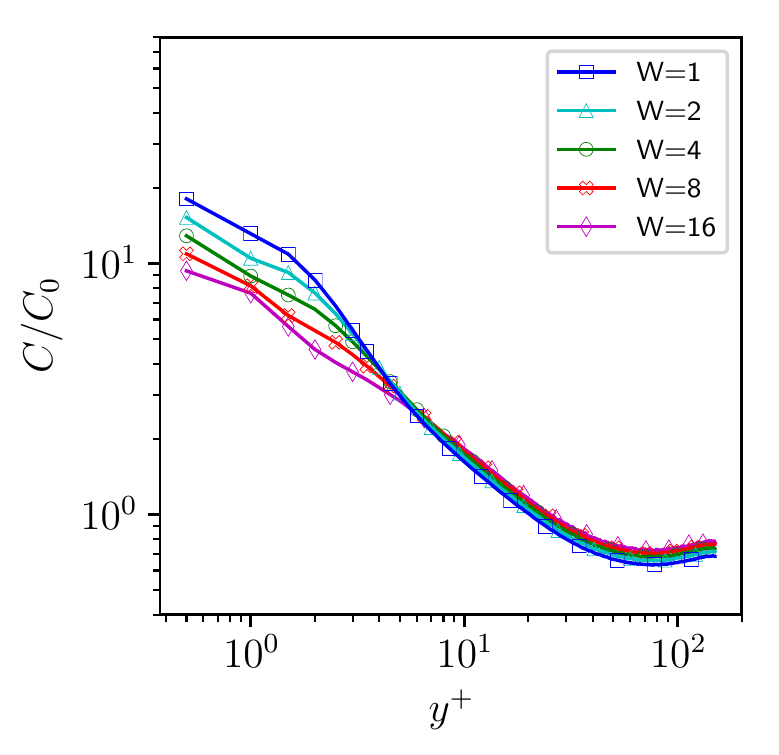}\\
	(a) \hspace{0.33\linewidth} (b) \hspace{0.33\linewidth} (c)
	\caption{Relative concentration profiles of $St^+ = 8$ particles at $\Phi_V = 1 \times 10^{-4}$, simulated with $N_c = N_p / W$ computational particles and (a) no correction, $d_c = d_p$; (b) low Stokes number correction, $h = 1$; and (c) high Stokes number correction, $h = 0$.}
	\label{fig:St8-h1}
\end{figure*}

Figures \ref{fig:St128-h0}-\ref{fig:St8-h1} show concentration profiles obtained from simulations with a reduced number of computational particles, indicated by increasing statistical weight, $W = N_p / N_c$, up to a factor of $16$. The order is from large to small Stokes number, beginning with $St^+ = 128$ in Figure \ref{fig:St128-h0} and decreasing to $St^+ = 8$ in Figure \ref{fig:St8-h1}. The Stokes number is changed by altering the particle-to-fluid density ratio. The bulk volume fraction, $\Phi_V$, is increased as the Stokes number is decreased to keep the mass fraction from changing dramatically, keeping the simulations in a range where two-way coupling does not strongly influence the concentration profiles.

The results for $St^+ = 128$ are shown in Figure \ref{fig:St128-h0}. In panel (a), the baseline result is obtained by making no correction to account for changes in particle-particle collisions. In that case, the sensitivity of the concentration profile to collision-driven changes in turbophoresis, as earlier demonstrated in Figure \ref{fig:wcollSt}, is apparent. Simply decreasing the number of computational particles without correction leads to a lower number of collisions per particle. As a result, the turbophoretic drift is stronger and larger concentrations are observed near the wall. In fact, with a $16$x factor reduction in number of computational particles, the near wall concentration is over-predicted by an order of magnitude compared to the $N_c = N_p$ case.

Panels (b) and (c) of Figure \ref{fig:St128-h0} show results using the enhanced collision radius model with $h = 1$ and $h = 0$, respectively. When $h = 1$, the discrepancies in predicted concentration profile with increasing $W$ are certainly less than in the case of no correction. Nevertheless, these discrepancies grow with increasing reduction in particle count, indicating the lack of statistical invariance with respect to changing $W$. In contrast, the concentration profiles using $h = 0$ approximately collapse even as $W$ is increased. The reason $h = 0$ succeeds in collapsing the concentration profiles is tied to the relatively large Stokes number. At $St^+ = 128$, the particles collide in a ballistic manner and therefore a larger correction of $d_c = d_p \sqrt{W}$ is necessary to maintain the same number of collisions per particle as the number of particles is reduced. The smaller correction of $d_c = d_p \sqrt[3]{W}$, being appropriate for smaller Stokes numbers, still under-predicts the number of collisions with reduced number of computational particles.

% % % % under the hood St = 128

%\begin{figure*}[t]
%	\includegraphics[width=0.335\linewidth]{coll_freq_Re150_St128_Phi1en5_h0.pdf}
%	\includegraphics[width=0.31\linewidth]{coll_w_avg_Re150_St128_Phi1en5_h0.pdf}
%	\includegraphics[width=0.335\linewidth]{vy_rms_Re150_St128_Phi1en5_h0.pdf}\\
%	(a) \hspace{0.33\linewidth} (b) \hspace{0.33\linewidth} (c)
%	\caption{For particle-laden channel flows with $St^+ = 128$ and $\Phi_V = 10^{-5}$, using $h = 0$, (a) particle-particle collision frequency; (b) average relative particle velocity at collision; and (c) root-mean-square of particle wall-normal velocity.}
%	\label{fig:St128-ext}
%\end{figure*}
%
%While the $h = 0$ scheme appears to work well for the concentration profiles at $St^+ = 128$, Figure \ref{fig:St128-ext} looks under the hood to confirm that accurate concentration profiles are obtained at $h = 0$ for consistent reasons. The collision rate as a function of wall-normal distance is shown in Figure \ref{fig:St128-ext}(a). Note that this collision rate is obtained by multiplying the raw number of collisions by $W$ in accordance with Eq.\ \eqref{eq:super-pdf-def}. Over most of the channel, the collision rate is invariant with $W$ when $h = 0$ is used. Some discrepancies can be seen very close to the wall. It is in this region, $y^+ \sim 1$, that the increasing finite size of the particles' collision radius makes a noticeable difference. This has a small influence on the concentration shown in Figure \ref{fig:St128-h0}(c), as evident from the slight divergence of the curves very close to the wall.

% % % %

Figure \ref{fig:St32-h0} shows the results for $St^+ = 32$. Again, with no correction, panel (a) shows that the concentration profile changes dramatically as the number of computational particles is reduced. The low Stokes number correction, $h = 1$, in panel (b) still shows significant increasing discrepancy with increasing $W$. In panel (c), the ballistic scale correction, $h = 0$, demonstrates excellent performance for $St^+ = 32$, as was also true for $St^+ = 128$. Again, the concentration profiles collapse for all values of $W$, signaling that invariance with respect to $W$ as been achieved. Thus, $St^+ =32$ and $St^+ = 128$ both appear to be safely in the high Stokes number regime anticipated in \S \ref{sec:super-size}.

However, continuing to decrease the Stokes number leads to different results. Figure \ref{fig:St8-h1} shows concentration profiles for simulations at $St^+ = 8$ for decreasing number of particle concentrations. Results similar to the higher Stokes numbers are obtained when no correction is performed, panel (a). However, judging from panels (b) and (c) of Figure \ref{fig:St8-h1}, the low Stokes number correction, $h = 1$, appears to lead to better results. This qualitative trend is exactly as expected from the model development in \S \ref{sec:super-size}. In the case of lower Stokes numbers, an increase in collision radius also increases the relative velocity with which the particles collide. As a result, the collision radius needs only to be scaled up as $d_c = d_p \sqrt[3]{W}$ to maintain accurate collisional activity with decreasing number of computational particles. This is reflected by the invariance of the concentration profile with respect to $W$ in Figure \ref{fig:St8-h1}(b). Panel (c) shows that the $h = 0$ scaling $d_c = d_p \sqrt{W}$ derived from high Stokes number assumptions actually over-corrects for collisional effects as $W$ is increased. In that case, the collision radius is made too large and the number of collisions actually increases with decreasing number of computational particles.

Taken together, the results shown in Figures \ref{fig:St128-h0}-\ref{fig:St8-h1} verify the expectations outlined for the large and small Stokes number limits, Eq. \eqref{eq:super-Holder-limits}. A simple scaling correction applied universally across all particles and all collisions works surprisingly well in generating collisional effects to maintain accurate concentration profiles even as the number of particles is dramatically reduces. Before, leaving this section, a few notes are in order. First, results at $St^+ = 512$ (not shown) indicate that the $h = 0$ correction continues to succeed at $St^+$ is further increased beyond what it shown here. At $St^+ = 2$, the decreasing slip velocity between the particle and surrounding fluid leads to a decrease in the turbophoretic drift, see Figure \ref{fig:Phi0}. As a result, concentration profiles at $St^+ = 2$ are not as sensitive to collisions in the volume fractions studied here. It follows, then, that even without a correction for the collisional radius, accurate concentration profiles can be obtained in the super-particle approach.

\subsection{Hybrid model \label{sec:hybrid-model}}
While the choice $h = 0$ seems to be effective for $St^+ \geq 32$ but $h = 1$ works better for $St^+ \leq 8$, general simulation scenarios are not necessarily characterized by a single Stokes number due to complex flow inhomogeneities and multidisperse particle size distributions encountered in real flows. It is more desirable to specify a unique $h$ value for each collision based on the properties of the collision known to the simulation. The basic difference between $h = 0$ and $h = 1$ is whether particles colliding at a separation $d_c$ would have a significantly different (lower) collision velocity if allowed to approach to a separation of $d_p$ before colliding. Therefore, it is straightforward to propose a dimensionless parameter,
\begin{equation}
b = \frac{|v_{rel}| \tau_p}{d_c - d_p} = \frac{\text{stopping distance}}{\text{additional distance to collision}},
\end{equation}
which encapsulates this behavior. The numerator is the stopping distance of a particle under Stokes drag, which indicates the relative distance the particles can travel before relaxing significantly toward local relative flow velocities. The denominator is simply a measure of the additional distance the particles must travel relative to each other before a collision based on their physical diameter. When $b$ is large, the particles respond very slowly to local relative flow near collision, indicating a mostly ballistic interaction ($h = 0$). In the opposite limit of small $b$, the particles adjust nearly instantaneously to the local flow and thus experience a collision in the `smooth' regime ($h = 1$). A simple model containing these asymptotics is
\begin{equation}
h = \exp\left( - \frac{b}{b_{tr}} \right),
\end{equation}
where $b_{tr} = 32$ is set empirically. This leads to an implicit formula for $h$, namely,
\begin{equation}
h =  \exp\left[ - \frac{|v_{rel}| \tau_p}{b_{tr} d_p} \left( W^{1/(2+h)} - 1 \right)^{-1} \right],
\end{equation}
but this is observed to converge quickly under iteration. Since $d_c$ is not known until after collision properties are computed, collision detection during each time step is done assuming the maximum $d_c$, namely, assuming $h = 0$ at first. Once $h$ is determined, the diameter and restitution coefficient of the collision are adjusted according to Eqs. \eqref{eq:super-diac} and \eqref{eq:super-ec}.

\begin{figure*}[t]
	\includegraphics[width=0.33\linewidth]{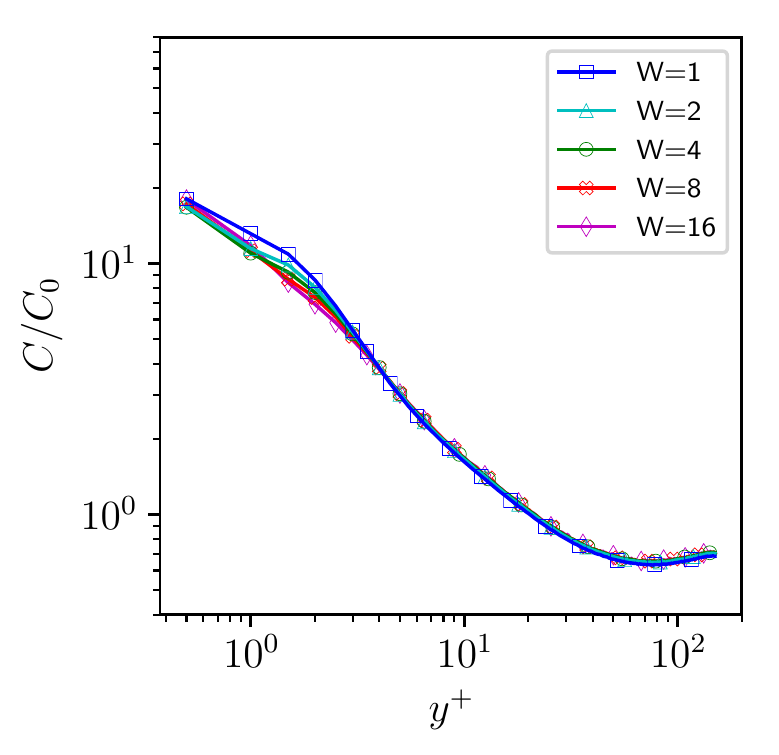}
	\includegraphics[width=0.33\linewidth]{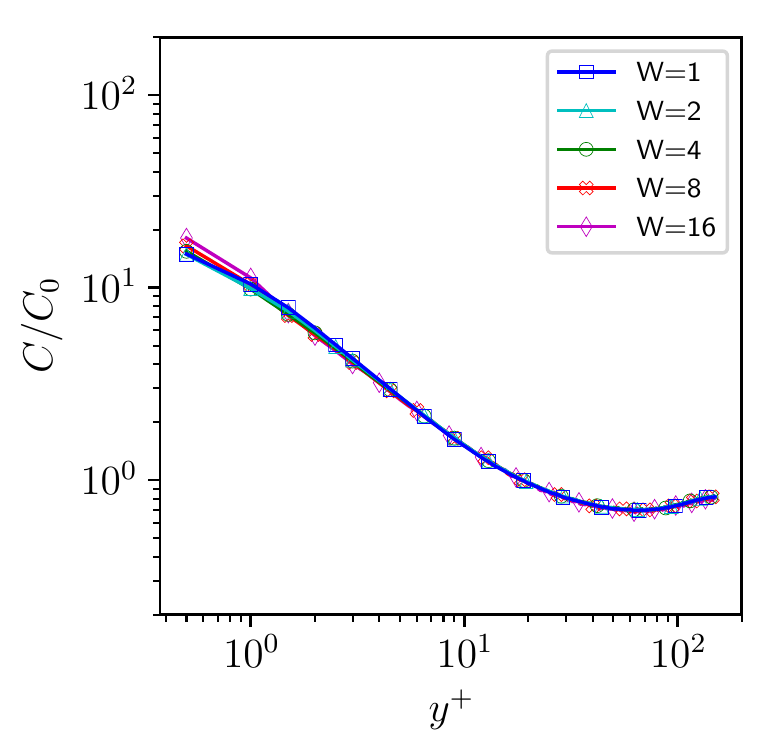}
	\includegraphics[width=0.33\linewidth]{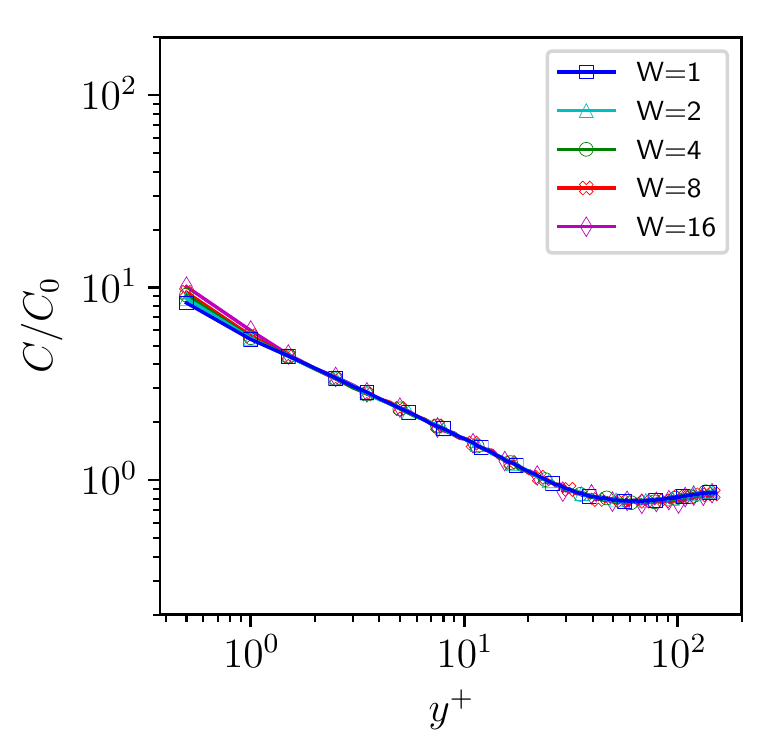}\\
	(a) \hspace{0.33\linewidth} (b) \hspace{0.33\linewidth} (c)
	\caption{Relative concentration profiles of $N_c = N_p / W$ super-particles at (a) $St^+ = 8$, $\Phi_V = 1 \times 10^{-4}$; (b) $St^+ = 32$, $\Phi_V = 3 \times 10^{-5}$; and (c) $St^+ = 128$, $\Phi_V = 1 \times 10^{-5}$.}
	\label{fig:StX-C-btr032}
\end{figure*}

\begin{figure*}[t]
	\includegraphics[width=0.33\linewidth]{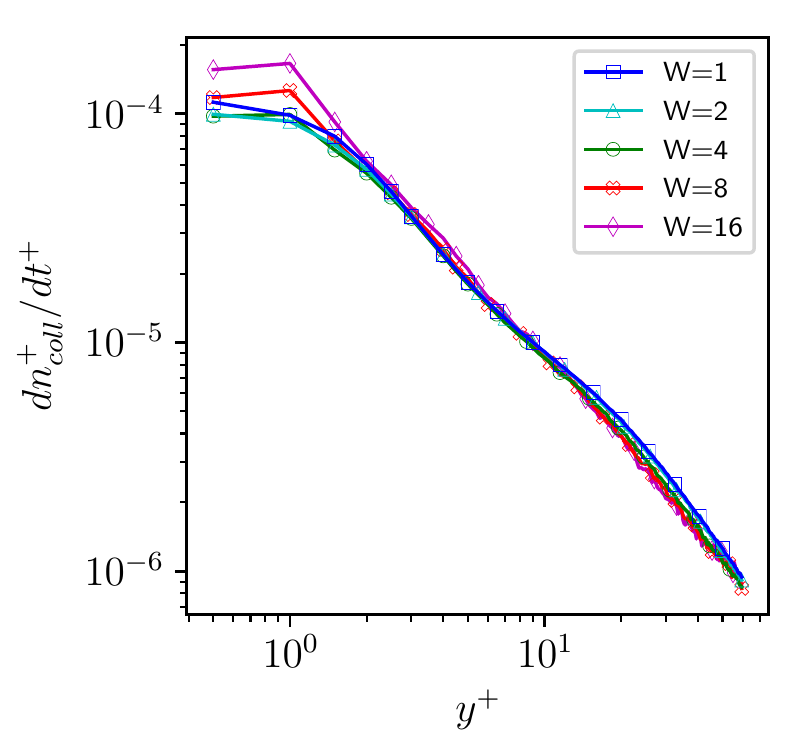}
	\includegraphics[width=0.33\linewidth]{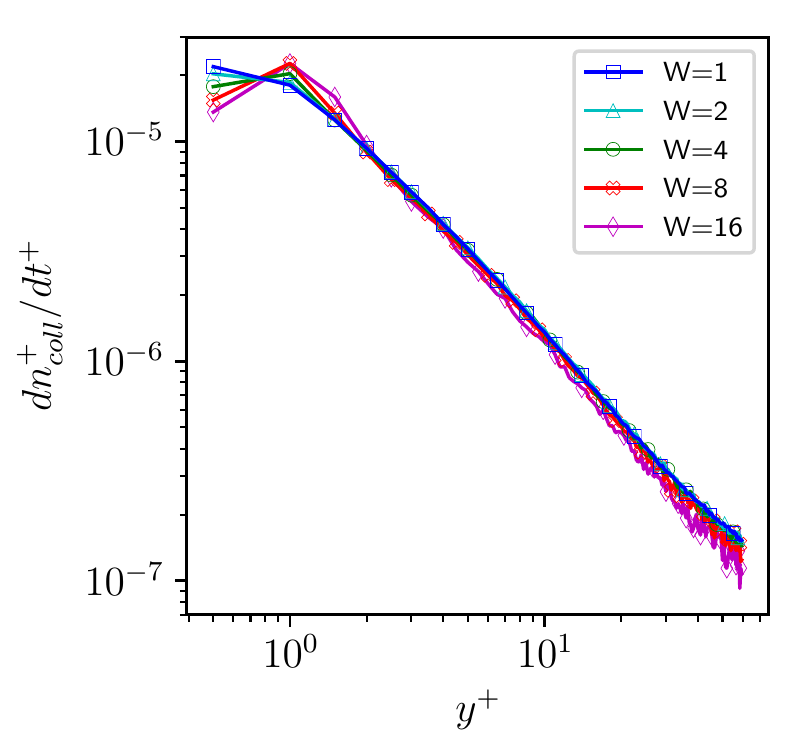}
	\includegraphics[width=0.33\linewidth]{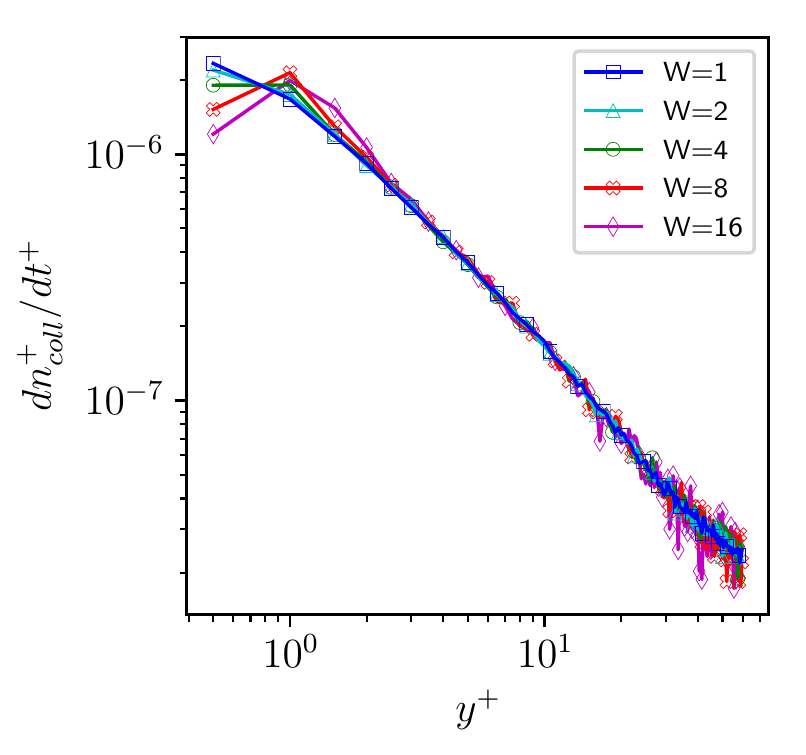}\\
	(a) \hspace{0.33\linewidth} (b) \hspace{0.33\linewidth} (c)
	\caption{Collision frequency as a function of wall-normal distance for $N_c = N_p / W$ super-particles using the hybrid model with $b_{tr} = 32$ at (a) $St^+ = 8$, $\Phi_V = 1 \times 10^{-4}$; (b) $St^+ = 32$, $\Phi_V = 3 \times 10^{-5}$; and (c) $St^+ = 128$, $\Phi_V = 1 \times 10^{-5}$.}
	\label{fig:StX-freq-btr032}
\end{figure*}

\begin{figure*}[t]
	\includegraphics[width=0.33\linewidth]{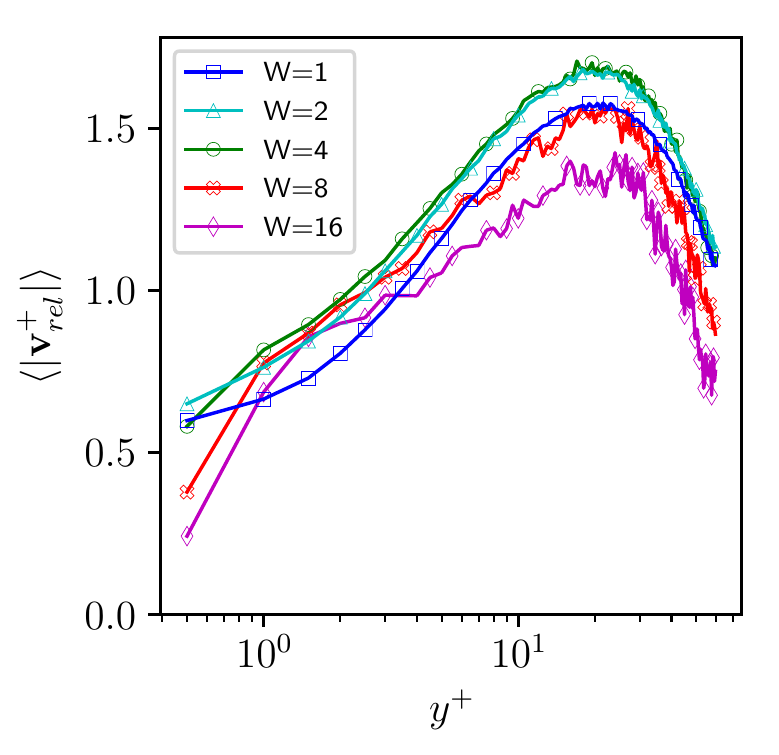}
	\includegraphics[width=0.33\linewidth]{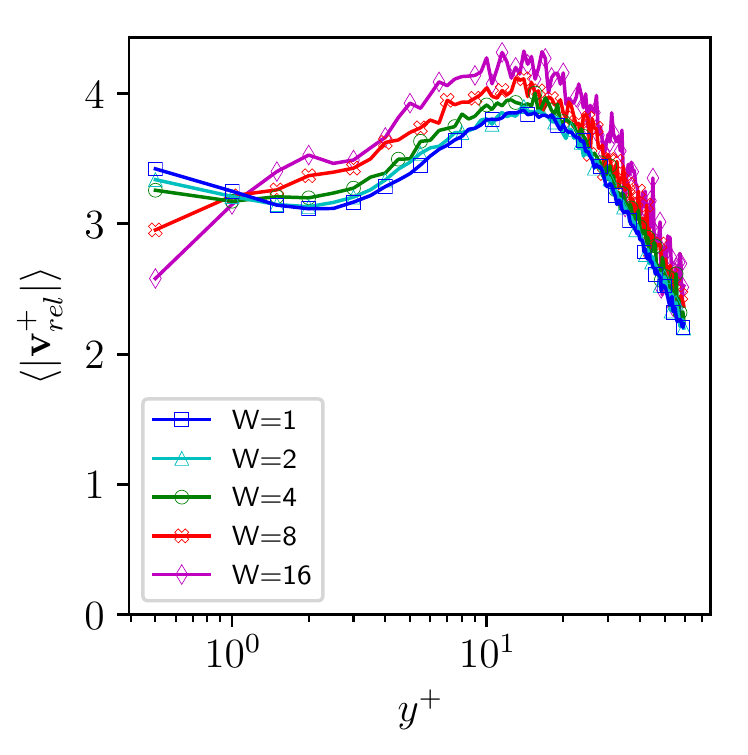}
	\includegraphics[width=0.33\linewidth]{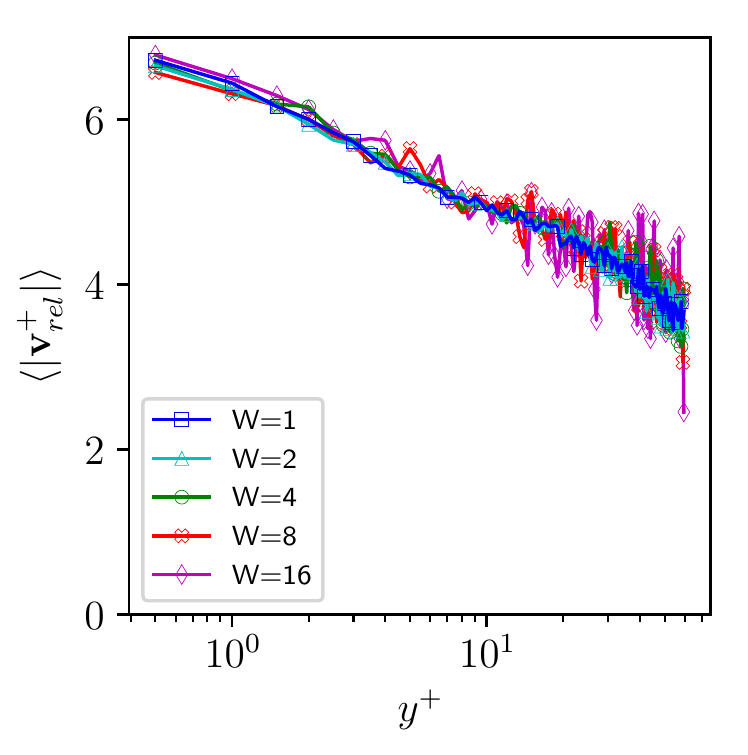}\\
	(a) \hspace{0.33\linewidth} (b) \hspace{0.33\linewidth} (c)
	\caption{Average relative velocity of particle-particle collisions as a function of wall-normal distance for $N_c = N_p / W$ super-particles using the hybrid model with $b_{tr} = 32$ at (a) $St^+ = 8$, $\Phi_V = 1 \times 10^{-4}$; (b) $St^+ = 32$, $\Phi_V = 3 \times 10^{-5}$; and (c) $St^+ = 128$, $\Phi_V = 1 \times 10^{-5}$.}
	\label{fig:StX-vrel-btr032}
\end{figure*}

The results of this super-particle collision model for the channel flow at $Re_* = 150$ are shown in Figure \ref{fig:StX-C-btr032} for three different Stokes numbers. It is apparent that, for each Stokes number, relatively good invariance is obtained with respect to $W$. This means that accurate concentration profiles are obtained even when the number of computational particles is dramatically reduced. The hybrid enhanced collision radius model, therefore, provides a realistic approach for the super-particle approach in situations where collisions are key to the accuracy of the simulation. To further support this assertion, additional quantities are shown in Figure \ref{fig:StX-freq-btr032} and \ref{fig:StX-vrel-btr032}. Note that wall-normal distances only up to $y^+ = 60$ are shown in these two figures due to the rarity of collisions, and hence noisy data further from the wall.

Figure \ref{fig:StX-freq-btr032} shows how collision frequency varies with distance from the wall for increasing statistical weight, $W$. The curves largely overlap for all three Stokes numbers, which is further evidence that particle statistics are roughly invariant with changes in $W$. Some discrepancy is noticeable within $y^+ \leq 1$, and this is reflected in Figure \ref{fig:StX-C-btr032} by some slight discrepancies in the near wall concentration. The discrepancy in collision rate, and hence concentration, in the very near wall region is due to the larger finite diameter relative to important flow features in the the viscous sublayer caused by the enhanced collision radius effect ($d_p^+ = 0.5$, so $d_c^+$ increases up to $2.0$ in the worst case shown). Although this effect remains minor for the cases shown here, this illustrates the main limitation of the enhanced collisional radius model developing in this paper. While the physical particle size may satisfy $d_p \lesssim \eta$ in a general turbulent flow, the distance at which super-particles collide may exceed this limitation if $W$ is taken too large. The collision radius for particle-wall collisions is unchanged in the super-particle approach, still computed based on the physical particle diameter. Tri-linear interpolation is used to obtain the fluid velocity at each particle location, even below the first grid point. Tests with higher-order interpolation showed no significant changes in the concentration profile \citep{Johnson2019}.

The average magnitude of relative velocity at collision is show in Figure \ref{fig:StX-vrel-btr032}. The noise in these plots serves as a reminder by illustrating that statistical convergence requires longer averaging times or more snapshots in the super-particle approach simply because less samples are available. Nonetheless, it can be seen that the super-particle approach with enhanced collision radius scaling performs reasonably well. The discrepancies, particularly for $y^+ \leq 10$ are larger than for concentration or collision frequency in the $St^+ = 8$ and $St^+ = 32$ cases. At higher $St^+ = 128$, the collapse with varying $W$ is more impressive. The lower Stokes number particles interact more nontrivially with near-wall coherent structures of the turbulent flow, which leads to less accuracy for the finer details in a super-particle approach.

It should be appreciated that the collision treatment here is not stochastic in nature. Each collision is computed in the same deterministic way as in the simulation with the full number of particles. While a hard sphere collision model was used in this study, the model is directly applicable to other approaches such as the soft sphere collision model. In principle, it could also be extended to a resolved particle approach as well, though some of the benefits of resolving the flow around individual particles would necessarily be forfeited, such as the capturing of the lubrication interaction near collision or accurately simulating particles for $d_p \gtrsim \eta$.

%\section{Application to PSAAP Duct}
%show that at high St or high volume fraction, very naive approaches are relatively accurate, Maxime's results

\section{Summary \& Conclusions \label{sec:conclusion}}

The simulation of particles subject to turbophoresis in a turbulent wall-bounded flows motivated the present work. At very low volume fractions, turbophoresis leads to dramatically enhanced concentrations in a thin layer of fluid adjacent to the wall. As volume fraction increases, particle-particle collisions play a key role in limiting the turbophoretic drift and reducing the near-wall peak in concentration. It was demonstrated here that this trend occurs in such a way that the collisions effectively create a maximum near-wall concentration that can be sustained by turbophoresis. Therefore, simply using a reduced number of computational particles can lead to large errors in near-wall concentration levels.

In this paper, an enhanced collision radius model was introduced to yield simulation results approximately invariant with respect to the number of computational particles used in the simulation. This treatment of simulated collisions for super-particles extends of the method of \citet{Garg2009}, see also \citet{Subramaniam2013} by artificially enhancing the collision radius of the particles, i.e. the separation distance at which two particles collide. The primary challenge in prescribing such a treatment is to specify how quickly the collision radius must grow as the number of computational particles decreases so as to  keep collision rates more or less constant.

%Because of the sensitivity of turbophoresis to volume fraction (via collisions), simply decreasing the number of computational particles in a simulation leads to significant changes in the concentration field and other particle statistics. 
%The particle statistics can be predicted with more accuracy using the enhanced collision radius treatment.
The specification of the collision radius can be done using scaling arguments for collision rates to approximate the conditions for statistical equivalence between the physical and computational particle ensembles, which is based on assigning a statistical weight to each computational particle. The proposed scalings of the collision radius with statistical weight were based on analytically established limits for low and high Stokes numbers. The method was first tested in these limits and showed promising results. Further, a hybrid model was designed to bridge these limits by relating the scaling of the collision radius to the properties of each individual collision. The hybrid model locally detects the extent to which each particle-particle collision is of a ballistic nature. As a result, the proposed model results in particle statistics that produce very similar particle statistics even as the number of computational particles is reduced.

\section*{Acknowledgements}
This investigation was funded by the Advanced Simulation and Computing program of the US Department of Energy's National Nuclear Security Administration via the PSAAP-II Center at Stanford, Grant No. DE-NA0002373. Jeremy Horwitz is thanked for a detailed review of this manuscript and for insightful discussions about the super-particle framework and his complementary research into two-way coupling effects. This work also benefited from interaction with Parviz Moin, Mahdi Esmaily, Ali Mani, Maxime Bassenne, Javier Urzay, and Lluis Jofre-Cruanyes. I would also like to thank Cristian Marchioli (University of Udine, Italy) for sharing benchmark numerical results that were useful in verifying our simulation code.

%\appendix
%\section*{References}

\bibliographystyle{elsarticle-harv}
\bibliography{turbophoresis_super.bib}

\end{document}